\documentclass[a4paper,11pt]{article}
\usepackage{aaskaiid}
\usepackage{orcidlink}
\title{Coherent and Incoherent Emission from the Ordered Magnetospheres of Low-Mass Stars, UCDs, and Massive Stars }
\ShortTitle{Ordered Magnetosphere Radio Stars}

\author[1]{Francesco Cavallaro\orcidlink{0000-0003-1856-6806}}
\ShortName{F. Cavallaro et al.} 
\author[1]{Paolo Leto\orcidlink{0000-0003-4864-2806}}
\author[2,3]{Barnali Das\orcidlink{0000-0001-8704-1822}}
\author[1]{Corrado Trigilio\orcidlink{0000-0002-1216-7831}}
\author[1]{Grazia Umana\orcidlink{0000-0002-6972-8388}}
\author[1]{Cristobal Bordiu\orcidlink{0000-0002-7703-0692}}
\author[1]{Filomena Bufano\orcidlink{0000-0002-3429-2481}}
\author[1]{Carla S. Buemi\orcidlink{0000-0002-7288-4613}}
\author[4,5]{Joseph R. Callingham\orcidlink{0000-0002-7167-1819}}
\author[6]{Laura Driessen\orcidlink{0000-0002-4405-3273}}
\author[1]{Adriano Ingallinera\orcidlink{0000-0002-3137-473X}}
\author[1]{Sara Loru\orcidlink{ 0000-0001-5126-1719}}
\author[7]{Stanley Owocki\orcidlink{0000-0002-0498-4735}}
\author[1]{Simone Riggi\orcidlink{ 0000-0001-6368-8330}}
\author[1]{Alan C. Ruggeri\orcidlink{ 0000-0002-1556-2474}}
\author[8]{Giovanni Sabatini\orcidlink{0000-0002-6428-9806}}
\author[9]{Matt E. Shultz\orcidlink{0000-0003-1387-5044}}
\author[10]{Alessio Traficante\orcidlink{0000-0003-1665-6402}}

\affiliation[1]{INAF–Osservatorio Astrofisico di Catania, Via S. Sofia 78, I-95123 Catania, Italy}
\emailAdd{francesco.cavallaro@inaf.it}
\affiliation[2]{National Centre for Radio Astrophysics, Tata Institute of Fundamental Research, Pune University Campus, Pune-411007, India}
\affiliation[3]{CSIRO, Space and Astronomy, PO Box 1130, Bentley, WA 6102, Australia}
\affiliation[4]{ASTRON, The Netherlands Institute for Radio Astronomy, Oude Hoogeveensedijk 4, Dwingeloo, 7991 PD, The Netherlands}
\affiliation[5]{Anton Pannenkoek Institute for Astronomy, University of Amsterdam, Science Park 904, 1098 XH, Amsterdam, The Netherlands}
\affiliation[6]{Sydney Institute for Astronomy, School of Physics A28, University of Sydney, NSW 2006, Australia}
\affiliation[7]{Department of Physics and Astronomy, University of Delaware, 217 Sharp Lab, Newark, Delaware, 19716, USA}
\affiliation[8]{INAF, Osservatorio Astrofisico di Arcetri, Largo E. Fermi 5, I-50125, Firenze}
\affiliation[9]{ESO - European Organisation for Astronomical Research in the Southern Hemisphere, Casilla 19001, Santiago 19, Chile}
\affiliation[10]{INAF-IAPS, Via Fosso del Cavaliere, 100, 00133 Rome, Italy}

\abstract{Massive early-type (B/A) stars and ultracool dwarfs (UCDs) represent two distinct regimes in which ordered, large-scale magnetospheres are observed. In rapidly rotating massive stars, incoherent radio emission is explained by the centrifugal breakout (CBO) mechanism: plasma confined within the rigidly rotating magnetosphere accumulates beyond the co-rotation radius, where centrifugal forces trigger breakout events and magnetic reconnection, generating non-thermal electrons that produce incoherent gyro-synchrotron emission. Empirically, the radio luminosity correlates with the power released by CBO events, establishing a clear link between stellar rotation, magnetic confinement, and radio output.

In UCDs, persistent non-thermal radio emission exhibits similar luminosity trends to those of massive magnetic stars, despite the absence of strong stellar winds. This similarity suggests that a CBO-like process may also operate in these fully convective, low-mass objects, though the plasma source and acceleration mechanisms remain uncertain. In both classes, coherent electron cyclotron maser emission (ECME), characterized by strong polarization and rotational modulation, is observed, indicating common magnetospheric processes analogous to planetary auroral emission.

The Square Kilometre Array (SKA) will be able to deeply observe about 70\% of the sky. We expect to observe $\sim 1000$ UCDs, enabling better statistical analysis of their emission and a test of the CBO hypothesis.

}

\begin{document}
\newcommand{\actaa}{Acta Astron.} 
\newcommand{\araa}{ARA\&A} 
\newcommand{\aar}{A\&ARv} 
\newcommand{\aapr}{A\&ARv} 
\newcommand{\ab}{Astrobiol.} 
\newcommand{\aj}{AJ} 
\newcommand{\apj}{ApJ} 
\newcommand{\apjl}{ApJL} 
\newcommand{\apjs}{ApJSS} 
\newcommand{\ao}{Appl. Opt.} 
\newcommand{\apss}{Astro. \& Space Sci.} 
\newcommand{\aap}{A\&A} 
\newcommand{\aaps}{A\&AS.} 
\newcommand{\baas}{Bull. Am. Astron. Soc.} 
\newcommand{\caa}{Chinese A\&A} 
\newcommand{\cjaa}{Chinese J. A\&A} 
\newcommand{\cqg}{Class. Quantum Gravity} 
\newcommand{\gal}{Galaxies} 
\newcommand{\gca}{Geo. Cosmo. Acta} 
\newcommand{\icarus}{Icarus} 
\newcommand{\jcap}{JCAP} 
\newcommand{\jgr}{J. Geophys. Res.} 
\newcommand{\jgrp}{J. Geophys. Res. Planets} 
\newcommand{\jqsrt}{J. Quant. Spectrosc. Radiat. Transf.} 
\newcommand{\memsai}{Mem. SAIt} 
\newcommand{\mnras}{MNRAS} 
\newcommand{\nat}{Nature} 
\newcommand{\nastro}{Nat. Astron.} 
\newcommand{\ncomms}{Nat. Commun.} 
\newcommand{\nphys}{Nat. Phys.} 
\newcommand{\na}{New Astron.} 
\newcommand{\nar}{New Astron. Rev.} 
\newcommand{\physrep}{Phys. Rep.} 
\newcommand{\pra}{Phys. Rev. A} 
\newcommand{\prb}{Phys. Rev. B} 
\newcommand{\prc}{Phys. Rev. C} 
\newcommand{\prd}{Phys. Rev. D} 
\newcommand{\pre}{Phys. Rev. E} 
\newcommand{\prx}{Phys. Rev. X} 
\newcommand{\prl}{Phys. Rev. Let.} 
\newcommand{\psj}{Planet. Sci. J.} 
\newcommand{\planss}{Planet. Space Sci.} 
\newcommand{\pnas}{Proc. Natl Acad. Sci. USA} 
\newcommand{\procspie}{Proc. SPIE} 
\newcommand{\pasa}{PASA} 
\newcommand{\pasj}{PASJ} 
\newcommand{\pasp}{PASP} 
\newcommand{\rmxaa}{RMXAA} 
\newcommand{\sci}{Science} 
\newcommand{\sciadv}{Sci. Adv.} 
\newcommand{\solphys}{Sol. Phys.} 
\newcommand{\sovast}{Soviet Ast.} 
\newcommand{\ssr}{Space Sci. Rev.} 
\newcommand{\uni}{Universe} 

\setlength{\bibsep}{0.0pt} 
\maketitle

\section{Introduction}

Magnetically active stars across the Hertzsprung-Russell diagram host large-scale structured magnetospheres that mediate the interaction between the stellar magnetic field, stellar winds, and ambient plasma. Among the large number of types of radio stars with radio emission related to the presence of magnetic fields, those showing time-stable radio emission are stars hosting large-scale well-ordered magnetospheres. These magnetospheres serve as efficient sites for a variety of radio emission processes, both incoherent and coherent, driven by energetic particles propagating along field lines. This kind of ordered magnetosphere is present in both early-type (B/A) stars and low-mass ultracool dwarfs (UCD), and both exhibit similarities in their radio behavior. In particular, rotational modulation of the incoherent radio emission (see for example \citealt{Leone1993,leto2020a} for early-type magnetic stars and \citealt{McLean2011,Llama2018} for UCDs) and time-stable periodic coherent radio pulses (see \citealt{trigilio2000} and \citealt{das2021} for the A0-type magnetic star CU~Virginis; see \citealt{Hallinan2007} and \citealp{Lynch2015} for the M8.5-type TVLM513-46).

\begin{figure}
    \centering
    \includegraphics[width=0.7\linewidth]{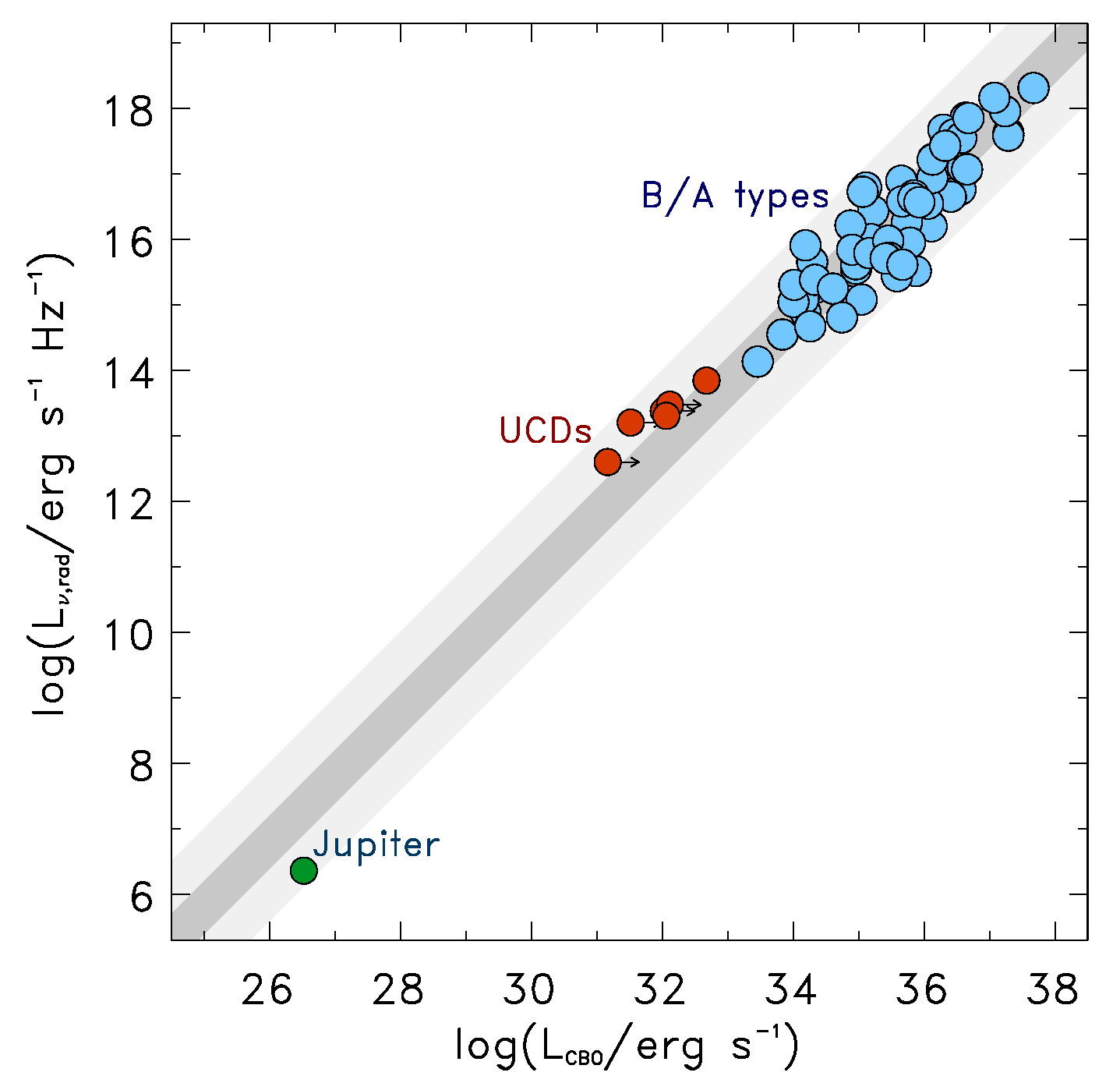}
    \caption{Relationship between CBO power $L_{\mathrm{CBO}}$ and radio luminosity $L_{\mathrm{\nu,rad}}$}
    \label{fig:1rel}
\end{figure}

Recently, it has been empirically discovered that the observed spectral radio luminosity ($L_{\nu, \mathrm{rad}}$) of B/A-type stars depends on a combination of fundamental stellar parameters, where the rotation speed plays a crucial role \citep{leto2021,shultz2022}. Due to the presence of the strong magnetic field dense plasma accumulates near the magnetic equatorial plane, at large distance centrifugal support leads to episodic centrifugal breakout (CBO) events (\citealt{uddoula2006}). In particular, the radio luminosity correlates strongly with the power released by centrifugal breakouts, \( L_{\mathrm{CBO}} \)  \citep{owocki2022}, as visible in Fig.\,\ref{fig:1rel}.

High-sensitive radio observations highlighted that UCDs may present incoherent emission due to gyro-synchrotron mechanism \citep{metodieva2017}, despite the apparent suppression of classical coronal magnetic activity \citep{mclean2012,williams2014}. Empirically, their incoherent radio luminosities appear to follow trends similar to those observed in B/A-type magnetic stars  (see Fig. \ref{fig:1rel}), ssuggesting the possible presence of related large-scale magnetospheric processes. Interestingly, the synchrotron radiation of the planet Jupiter \citep{depater2003} seems to satisfy the same scaling relationship found by studying the B/A type magnetic stars (see Fig. \ref{fig:1rel}). 
The radio emission from the magnetized planets of the solar system satisfies the generalized ``Radio Bode's law'', that is possibly extended also to the exoplanets case \citep{Zarka2007pss}. Bode's law correlates the power released by radio emission to the power released by kinetic or magnetic interaction of an external body with the magnetosphere. Possibly, the existence of a general rotational and magnetic dependence seen in the magnetospheric radio emission of stars surrounded by large-scale and stable magnetospheres suggests a common physical underpinning.

In parallel, a coherent radio emission component—arising from the electron cyclotron maser (ECM) instability—has been detected in both early-type stars \citep{trigilio2000,das2025b} and UCDs \citep{Berger2002,Guirado2025}. 
These emissions are highly polarized, beamed, and often rotationally modulated, forming a “radio lighthouse” \citep{trigilio2011}. In UCDs and very-low-mass stars, such coherent bursts occur despite strong suppression of coronal X-ray emission, challenging classical magnetospheric models \citep{williams2014}. The source of plasma in these magnetospheres is likely non-stellar, with proposed origins including moon–planet interactions analogous to the Jupiter–Io system \citep{Badman2015}.

Collectively, these findings demonstrate that the presence of structured magnetospheres and their interaction with plasma reservoirs—whether wind-fed or externally sourced—governs both incoherent and coherent emission processes. The Square Kilometre Array (SKA), with its sensitivity and spectral coverage, could advance our understanding of these mechanisms across the stellar mass spectrum. In particular, confirm or disprove the CBO model for UCDs and, if confirmed, understand what could cause the presence of a dense enough plasma around them. This paper reviews the current status of radio emission from ordered magnetospheres in massive stars and UCDs, and outlines the SKA's transformative potential for this field.

This chapter aims to provide the following take-home points for the SKA community:
\begin{enumerate}
    \item Observational status: A synthesis of coherent (ECME) and incoherent (gyro-synchrotron) emission from ordered magnetospheres in hot magnetic stars and ultracool dwarfs (UCDs);
    \item physical interpretation: A critical assessment of the centrifugal breakout (CBO) paradigm and its possible extension to UCDs;
    \item SKA discriminants: Identification of specific observational tests enabled uniquely by SKA sensitivity, bandwidth, and survey scale;
    \item forecast framework: Transparent and reusable estimates for detection yields in wide-area surveys and targeted observations, with clearly stated assumptions.
\end{enumerate}

\section{Coherent Emission Processes}  
\label{sec:coherent}

Coherent radio emission from main-sequence stars and sub-stellar objects can arise through two main mechanisms: (a) plasma emission and (b) electron cyclotron maser emission (ECME). In the context of emission produced within large-scale, ordered magnetospheres, ECME is generally the dominant process, responsible for coherent radiation from stars across the H-R diagram, brown dwarfs, and even planets. In this section, we focus on the stable form of ECME generated in long-lived magnetospheres, typically observed as periodic radio pulses rather than sporadic flares. For a broader picture about transient phenomena in radio stars please see \citet{Driessen.1.2026.SKA}.

The ECME mechanism produces highly beamed radiation, confined within a hollow cone aligned with the local magnetic field lines of each emitting region \citep{Melrose1982}. It is intrinsically narrow-band, with the emission frequency directly proportional to the magnetic field strength at the source \citep[e.g.][]{Treumann2006}. Consequently, different frequencies originate at different heights within the stellar magnetosphere. Drawing an analogy with auroral radio emission from magnetized planets in the solar system \citep{Zarka1998}, each frequency corresponds to a thin auroral ring located above the magnetic poles. 

Because ECME is a directional process, the emission becomes detectable only when the beam intersects the observer’s line of sight as the star rotates—producing the characteristic periodic pulses often compared to a “radio lighthouse” effect. 

In the following section, we summarize the main observational properties of ECME as identified in stellar systems where this phenomenon has been most extensively studied.

\subsection{ECM emission from hot magnetic stars}\label{subsec:ecme_hms}
Magnetic hot stars are characterized by their large-scale, kG-strength surface magnetic fields that are stable throughout their lifetimes. In most cases, these fields can be described as dipoles inclined to the stellar rotation axes \citep[e.g.][]{shultz2018}, in accordance with the oblique rotator model \citep[ORM;][]{Babcock1949,Stibbs1950}. Among the different types of stars and substellar objects that produce coherent radio emission, magnetic hot stars exhibit potentially the most stable form of this phenomenon, because of the extraordinary stability of their magnetic fields as well as the stable supply within the stellar magnetosphere of particles with unstable energy distribution powering the maser emission. Magnetic hot stars producing ECME are now called ``Main-sequence Radio Pulse emitters'' \citep[``MRPs'',][]{das2021}.

\begin{figure}
    \centering
    \includegraphics[width=0.7\linewidth]{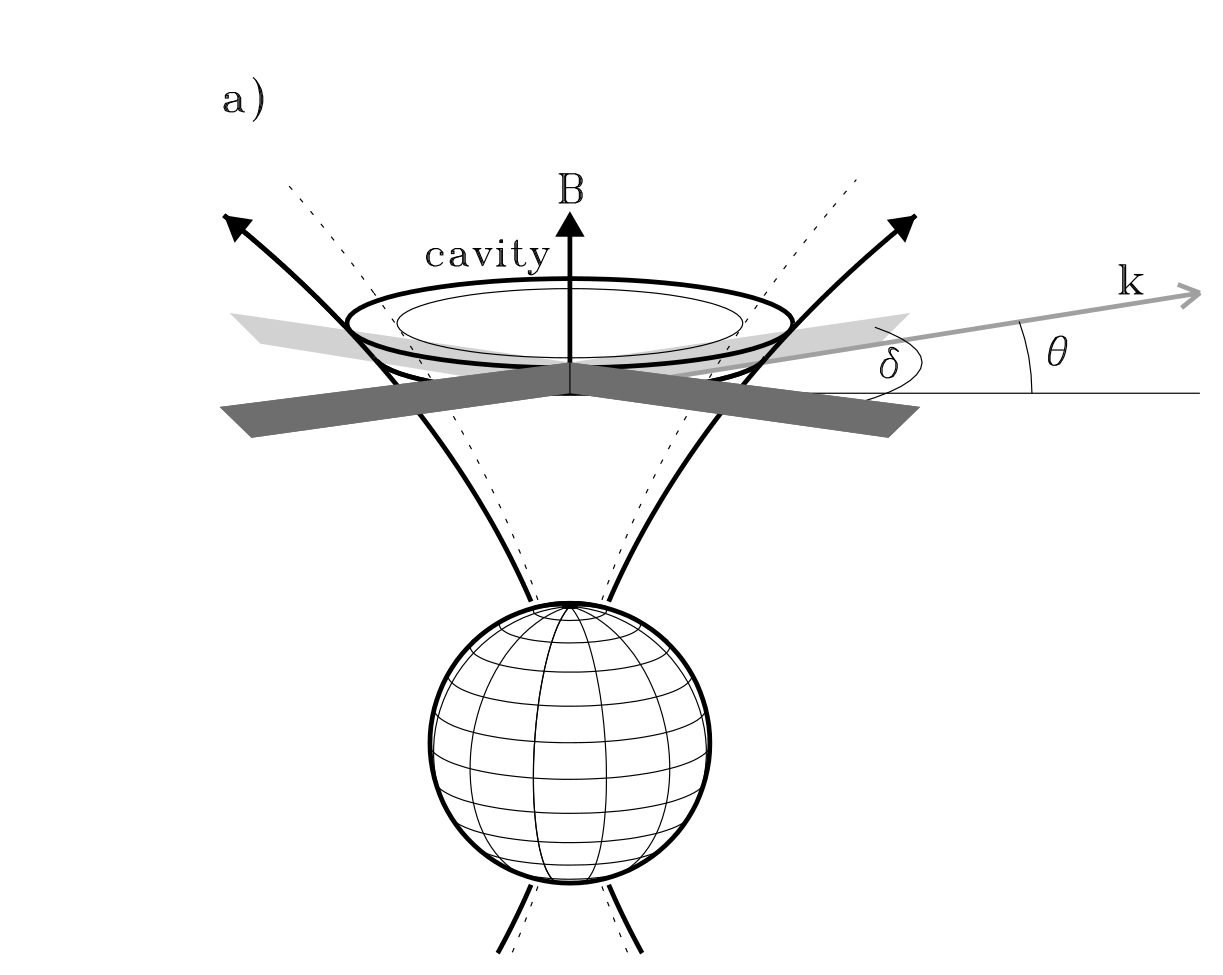}
\caption{Tangent plane beaming model. Figure taken from \citet{leto2016}.}
\label{tp}
\end{figure}

The first MRP was discovered by \citet{trigilio2000}. This star, CU\,Virginis (hereafter CU\,Vir), was found to produce two $\approx 100\%$ circularly polarized pulses at 1.4 GHz that are observable over a set of narrow rotational phase ranges surrounding the phases where the stellar longitudinal magnetic field $\langle B_\mathrm{z}\rangle$ is zero (called magnetic nulls). The combination of high circular polarization, high brightness temperature and their periodic arrivals around the magnetic nulls led \citet{trigilio2000} to attribute the emission to ECME. To explain the arrival of the pulses around magnetic nulls and their narrow width (duty cycle of $\approx 10\%$ for each pulse), \citet{trigilio2011} proposed the `tangent plane beaming model' inspired from the observed beaming geometry of auroral kilometric radiation from Earth. According to this model, radiation is emitted tangentially to the auroral rings (see Fig.\,\ref{tp}). This is obtained as sum of the individual ECME elementary sources radiating within the `hollow-cone beams' that are able to cross the line of sight. In addition to the new beaming model, \citet{trigilio2011} also pointed out the role of the magnetospheric plasma in determining the relative arrival times of pulses at different frequencies, which is subsequently investigated in much greater detail by \citet{leto2016} and \citet{das2020}.

The current MRP sample size is 22 \citep{trigilio2000,chandra2015,das2018,lenc2018,leto2019,das2019a,das2019b,leto2020a,leto2020b,pritchard2021,das2022a,das2022c,biswas2025,das2025b}. It is worth noting that seven of the sample were first discovered as candidates in surveys using the SKA pathfinders and precursors \citep[uGMRT, MWA and ASKAP;][]{chandra2015,lenc2018,pritchard2021,das2025b}. 
Among the 22 MRPs, nine were observed over wide enough frequency range to constrain the upper cut-off frequencies \citep[e.g., see Table 1 of][and references therein]{das2022b}. The significance of the ECME upper cut-off is that it is expected to be related to the maximum magnetic field strength in the stellar magnetosphere, which for a star with dipolar magnetic field, is the magnetic field at the poles. This is indeed observed for ECME produced by Jupiter \citep{zarka2004}. However, for the MRPs, the ECME upper cut-off frequencies are found to be significantly smaller than the electron gyrofrequencies corresponding to the polar magnetic field strengths (obtained from spectropolarimetry) in all cases \citep{das2022b}. The underlying reason is not known. This is an important problem to resolve as it impacts the usage of ECME to estimate magnetic field strength in objects for which alternative magnetic measurements are much more challenging.

Another new phenomenon that came to light with ultra wideband observation of ECME is the strong frequency dependence of the `duty cycle'. For the first discovered MRP CU Vir, ECME is observable for $\approx$20\% of the rotation cycle at 1.5 GHz \citep[e.g.][]{trigilio2011}. This increases to nearly 40\% at 700 MHz \citep{das2021}. In addition, at 400 MHz, `off-pulse' emission (for which there are no higher frequency counterparts) were also observed \citep{das2021}. Most recently, off-pulse emission (secondary pulses) were also reported from another MRP HD 142990 at $\approx 1$ GHz showing that this is not a low-frequency only phenomenon \citep{das2025a}.  This is an important observation in the context of searching for ECME signatures not only to find new MRPs, but also for areas like star-planet interaction. Once fully characterized, this could help us in deciding optimal frequency ranges and/or times of observations to detect signatures of star-planet interaction (in the form of ECME). 

It has been proposed that strong frequency dependence of ECME properties could arise due to propagation effects in the magnetospheres of stars with highly misaligned rotation and magnetic axes \citep[both CU Vir and HD 142990 belong to this category,][]{das2020}. In those cases, the density distribution in the stellar magnetosphere is expected to be highly azimuthally asymmetric w.r.t. the magnetic axis. As a result, ECME at different frequencies that are produced at different heights from the stellar surface can experience very different plasma densities on their way to the observer. These effects can significantly impact the duty cycle, arrival times, and even the cut-offs. Although the propagation effects seemingly complicate the observable properties of ECME, it also means that ECME properties over wideband can be used to map the magnetospheric plasma density. This has recently been demonstrated by \citet{das2024}. 

To summarize, the study of ECME from MRPs has led to a number of new insights about the phenomenon that are also highly relevant for other magnetic systems. The current sample is, however, inadequate to help us fully exploit the potential of ECME as a stellar (sub-stellar) magnetospheric probe. The biggest limitation is the small sample that does not allow drawing robust conclusions. Besides, the current sample is dominated by MRPs discovered from targeted searches that involved imposing selection criteria based on pre-existing notions about what types of magnetic massive stars emit ECME. This is the reason that all but one MRPs are rapid rotators ($P_\mathrm{rot} <2$ days, the only exception is HD 79158 that has a period of $\approx 4$ days), since longer Prot means larger telescope time required to cover a given range of rotational phases. In contrast, the $P_\mathrm{rot}$ range of known radio-bright magnetic massive stars extend to $>10$ days \citep{Leto2025}. As all the MRPs have very similar $P_\mathrm{rot}$, it has not been possible to investigate the role of $P_\mathrm{rot}$ in driving ECME \citep[rotation has been established to play a key role for incoherent radio emission,][]{leto2021,shultz2022}. Similarly, to confirm or rule out magnetospheric propagation effects as the reason for seemingly anomalous behaviour of ECME as a function of frequencies, it will be important to observe a larger number of MRPs over wideband, preferably simultaneously.

\subsection{Advances in studying ECME from hot stars using the SKA}\label{subsec:ska_for_ecm}
With the combination of unprecedented sensitivity and resolution, the SKA (Mid and Low) will help us overcome this limitation. The detection statistics of the existing MRP sample suggests that the Band 1 (0.35-1.05 GHz) of the SKA-Mid will be a very suitable frequency range to search for new MRPs. The current sample spans a distance range up to $\approx 400$ pc (HD\,37017). With the SKA, we will be able to detect MRPs from a few kilo parsecs \citep[based on the observed spectral ECME luminosities of the existing sample,][]{das2022c} leading to a significant expansion of the MRP sample. Besides, the high sensitivity will enable observing at multiple bands using subarray mode. With the SKA-Mid alone, it will be possible to cover a frequency range of 350 MHz to 1.8 GHz simultaneously (Band 1 plus Band 2), which is wider than the frequency range of observations of most MRPs at present. The SKA-Low, on the other hand, will provide access to a new phase space. Little is known about the MRP properties below 400 MHz, though there are strong evidences suggesting that the ECME spectra from MRPs extend well below 400 MHz \citep[e.g.][etc.]{lenc2018,das2021}. With co-ordinated observations between the SKA-Low and SKA-Mid, one will be able to cover the frequency range of 0.05--1.8 GHz simultaneously, which will be pivotal in fully characterizing the phenomenon and thus enabling implementation of ECME as a versatile magnetospheric probe.

\section{Incoherent Emission Processes} 
\label{sec:incoherent}

The study of stellar radio emission has gained growing attention in recent years \citep{Matthews2025}. The mechanisms responsible for this emission involve both thermal and non-thermal processes \citep{Dulk1985}. In particular, the non-thermal component requires the presence of a magnetic field and a population of energetic electrons. For mildly relativistic electrons, the dominant emission process is gyro-synchrotron radiation, which produces a broad-band radio spectrum that is typically unpolarized or only weakly circularly polarized.

Stars possessing convective envelopes beneath their photospheres—mainly those of spectral type later than F—exhibit coronal magnetic activity and often show flare-like non-thermal radio emission. In contrast, stars of spectral type earlier than A, when radio emitters, display time-stable emission in both their incoherent \citep{Leto2012} and coherent \citep{trigilio2011} components.

In the following subsection, we summarize recent results obtained from the study of incoherent radio emission in early-type magnetic stars.

\subsection{The stable gyro-synchrotron radio emission from the hot magnetic stars}\label{subsec:incoherent_hms}
Incoherent radio emission from magnetized stars is dominated by the gyrosynchrotron
mechanism, powered by populations of mildly relativistic electrons spiraling in large-scale
magnetic fields. Unlike coherent processes, these emissions are broadband, weakly
polarized, and trace the global structure of the stellar magnetosphere. Radio observations across
the mass spectrum --- from early-type (B/A) magnetic stars to ultracool dwarfs
(UCDs) --- evidence an overall commonality in their radio behavior (see Fig.\,\ref{fig:1rel}) which reveals common physical drivers. 

In magnetic B/A type stars, large-scale kilogauss dipolar fields \citep{shultz2019,sikora2019a} confine the line-driven stellar wind into rigidly rotating magnetospheres (RRMs; \citealt{townsend2005}), enabling the accumulation of dense plasma in the magnetic equatorial plane. The outward centrifugal force acting on this co-rotating material can exceed the magnetic tension at distances beyond the Kepler co-rotation radius \( R_K \), resulting in centrifugal breakout (CBO) events \citep{uddoula2006}. These breakouts are hypothesized to generate magnetic reconnection in current sheets, leading to the production of non-thermal electrons \citep{hoshino2001,zweibel2009}. These semi-relativistic particles emit via the incoherent gyro-synchrotron mechanism, producing a broad-band radio spectrum modulated by stellar rotation \citep{trigilio2004,leto2006,leto2017}.

The resulting radio emission arises from dipole-like magnetic shell where the energetic electrons propagate, similar to radiation belts covering a large range of magnetic latitudes. The observing radio emission features of the incoherent radio emission from the BA-type magnetic stars is the almost flat radio spectrum, covering a wide spectral range from about hundred of MHz up to tens GHz due to the radial dependence of the magnetic field strength of the stellar magnetosphere region where the radio emission at a given radio frequency mainly arises,  and the rotational modulation, mostly given by the consequence of the changing optical depth across the magnetosphere whose orientation respect to the line of sight changes as function of the rotation cycle. In addition, the collection of a large number of radio measurements of BA-type magnetic stars with well known stellar parameters has demonstrated a robust correlation between radio luminosity and fundamental stellar parameters (\citealt{leto2021,shultz2022}). Specifically, the emitted
power correlates with the centrifugal breakout luminosity,
\begin{equation}
    L_{\mathrm{rad}} \propto L_{\mathrm{CBO}} \propto
    B_p^{2} R_\ast^{4.5} M_\ast^{-0.5} P_{\mathrm{rot}}^{-2},
    \label{eq:lrad_lcbo}
\end{equation}
as derived by \citet{owocki2022}. The inferred efficiency of conversion from centrifugal breakout power into non-thermal radio emission is extremely low ($\sim 10^{-19}$), but appears sufficient to account for the observed radio luminosities of currently known systems. The radio emission from magnetic B/A stars should also be considered within the broader context of feedback from massive stars, whose winds and magnetized outflows affect both their immediate surroundings and the larger Galactic environment \citep{Roshi.1.2026.SKA}.

\subsection{The use of SKA for indirect magnetometry of hot magnetic stars}\label{subsec:ska_indirect_magnetometry}

Reliable measurements of the magnetic fields of the early-type magnetic stars
have been performed only for stars located within 1-2 kpc from the Sun \citep{Shultz2019MNRAS485}. The Ap/Bp stars are uniformly distributed in space \citep{Renson2009} and approximately 10\% of them are magnetic stars \citep{Grunhut2012,Wade2016,sikora2019a}.

As previously discussed (see Sec.\,\ref{subsec:ska_for_ecm}), at the lower frequencies (namely the Band 1 and Band 2 of SKA), the radio spectra of the hot-magnetic star may be contaminated by the presence of the coherent radio emission component. In fact, multi-frequency radio observations of
early-type magnetic stars covering their entire rotational periods never reported highly polarized pulses of auroral origin at frequencies higher than 5\,GHz \citep{das2021}.
Therefore, the optimal radio frequency window to search for evidence of incoherent non-thermal radio emission from the early-type magnetic stars is Band 5. In particular, Band 5a, tuned at 6.6\,GHz, maximizes the observing sensitivity, allowing to reach the noise of about 0.5\,$\mu$Jy/beam in 1 hour of integration time, estimated using the SKAO Sensitivity Calculator for the AA4 configuration. In this case, the $3\sigma$ detection threshold is about 1.5\,$\mu$Jy/beam.

Looking at Fig.\,\ref{fig:1rel}, the average spectral radio luminosity of the B/A-type magnetic stars is roughly centered at $10^{16}$\,erg\,s$^{-1}$\,Hz$^{-1}$. Assuming such radio luminosity, SKA will be able to detect, with 1 hour integration time, all the hot-magnetic stars brighter than this average value up to a distance from the Sun of about 2\,kpc. All the stars brighter than $10^{17}$\,erg\,s$^{-1}$\,Hz$^{-1}$ will be instead visible up to the distance of the center of the Galaxy ($\approx 8$\,kpc). Whereas the few stars brighter than $10^{18}$\,erg\,s$^{-1}$\,Hz$^{-1}$ should be detectable wherever they are located in the Galaxy. 

Taking into account the relation between the radio luminosities of the stars and their fundamental stellar parameters, summarized by Eq.\,\ref{eq:lrad_lcbo}, the detection of the incoherent non-thermal radio emission from such kind of stars could be used as a useful tool for the estimation of the fundamental stellar parameters. In particular, the detection of the incoherent radio emission from B/A-type stars should be highly valuable for the indirect estimate of the magnetic field strength. Therefore, the SKA will be able to deeply study the magnetism of such kind of stars. In fact, SKA will be able to detect the radio emission from far B/A stars whose magnetic fields are not detectable by the current ground-based instruments  (i.e., for stars more distant than 2 kpc).

\section{The case of the Ultra Cool Dwarfs}\label{subsec:incoherent_ucd}

Strong magnetic fields, with intensities of several kilogauss, have been confirmed in very cool stars (spectral type later than M7) located at the bottom of the main sequence \citep{reiners2007}. These Ultra Cool Dwarfs (UCDs) are fully convective and therefore cannot sustain the classical dynamo mechanism, which operates in the separation region between the radiative internal stellar layers and the convective envelope, and which is responsible for the solar-like coronal magnetic activity.
Both theoretical \citep{Yadav2015} and observational studies \citep{Donati2006,Morin2010} support the presence of large-scale, well-ordered (mostly dipolar) magnetic fields in such fully convective objects.

Persistent, unpolarized radio emission has been detected in a subset of UCDs \citep{metodieva2017}, despite their weak coronal activity and low flaring rates \citep{mclean2012,williams2014}. This steady emission has a non-thermal origin and is attributed to incoherent gyrosynchrotron radiation. Its presence suggests that steady plasma processes occurring within their ordered magnetospheres --- rather than transient flare-driven events --- maintain the energetic electron populations. Empirically, the incoherent radio emission of UCDs shows a striking similarity to that observed in B/A-type magnetic stars, as illustrated by the position in the $L_{\nu,\text{rad}}$--$L_{\text{CBO}}$ diagram of the few UCDs having the incoherent radio emission component detected and also having reliable information of their stellar parameters (see Fig.~\ref{eq:lrad_lcbo}). The radio luminosities and stability of UCDs follow the same trends observed in massive magnetic stars, reinforcing the hypothesis of a shared underlying mechanism.  
The existence of large-scale, stable magnetic fields may therefore explain the continuity of magnetospheric behaviour across such widely different stellar regimes.

The incoherent emission from UCDs can also be modulated by stellar rotation \citep{McLean2011}. In hot magnetic stars, such modulation arises from a combination of frequency-dependent absorption by thermal plasma trapped along closed magnetic field lines and changes in the optical depth of the non-thermal gyrosynchrotron-emitting regions \citep{trigilio2004}. Similarly, in UCDs, the modulation may result from large-scale magnetospheric structures containing mildly relativistic electrons responsible for the non-thermal emission, together with anisotropies in the thermal plasma distribution. These anisotropies could be influenced by external sources, such as planetary companions or residual chromospheric outflows, which introduce absorption effects across the magnetosphere.

Population studies indicate that approximately 15\% of UCDs produce persistent incoherent radio emission \citep{kao2024}, which combined with the space density estimates from \citet{best2024} and \citet{cruz2007}, corresponds to about 12,500 detectable sources within 100~pc . These systems provide an ideal laboratory to test the applicability of the centrifugal breakout (CBO) paradigm beyond the main sequence.

The comparison across spectral types highlights a continuity in the underlying physical processes, with the main differences arising from the origin of the plasma supply, stellar winds in massive stars, versus external or chromospheric sources in UCDs.

\begin{figure}
    \centering
    \includegraphics[width=1\linewidth]{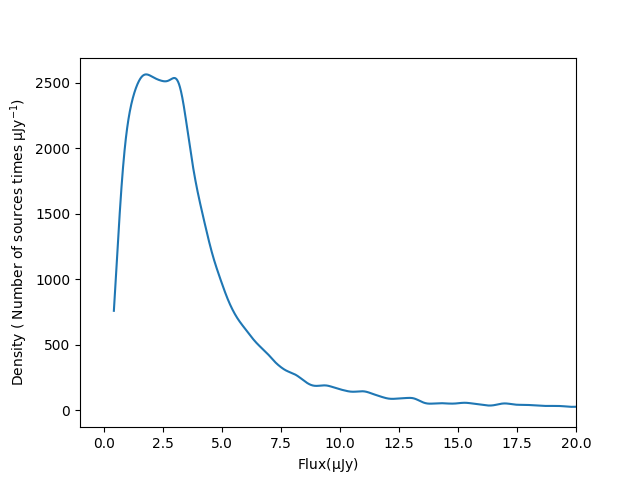}
    \caption{Differential source counts dN/dS of UCDs within a 100 pc radius sphere as a function of flux density.}
    \label{fig:3distri}
\end{figure}
In magnetic early type stars, centrifugal breakout (CBO) operates within wind-fed magnetospheres: radiatively driven stellar winds continuously supply plasma, which accumulates beyond the Kepler co-rotation radius until centrifugal stress triggers reconnection events. In this context, the existence of a plasma reservoir is not in doubt.

In UCDs, however, the situation is fundamentally different. Radiatively driven winds are expected to be extremely weak, and the coolest atmospheres are largely neutral, potentially suppressing sustained coronal outflows. A CBO-like interpretation in UCDs therefore requires a mechanism to load the magnetosphere with sufficient plasma and replenish it at at least the same rate of the particle loss rate.

At present, the plasma supply mechanism in UCDs remains uncertain. Plausible contributors include:

\begin{itemize}
    \item weak but non-zero outflows,
    \item satellite or star–planet interactions (Jupiter–Io analogues),
    \item sputtering or impact ionisation from close-in bodies or debris,
    \item reconnection-driven interchange processes that recycle magnetospheric plasma.
\end{itemize}

Without a clearly identified supply channel, the transplantation of the hot-star CBO paradigm to UCDs remains suggestive rather than demonstrated. One of the key roles of SKA observations will be to constrain whether a sustained plasma reservoir is present and dynamically replenished.
\subsection{SKA Discriminants of CBO-like Behaviour in UCDs}
The sensitivity and spectral coverage of the SKA will make it possible to test whether UCD magnetospheres are in a steady state, analogous to continuously occurring centrifugal breakout events, which explain the time-stable radio emission observed in magnetic hot stars, or undergo loading–release cycles.

Possible observational discriminants include:

\begin{itemize}
    \item Abrupt changes in quiescent continuum level or spectral turnover frequency followed by slower recovery, consistent with large-scale reconnection or magnetospheric emptying events;
    \item Systematic wideband spectral evolution associated with changes in magnetospheric plasma density or optical depth;
    \item Correlated variability between coherent ECME activity and the incoherent gyro-synchrotron component, indicative of a shared particle reservoir;
    \item Population-level scaling relations linking radio luminosity with stellar rotation and large-scale magnetic field properties.
\end{itemize}

Such observations will help determine whether UCD radio emission is primarily driven by CBO-like magnetospheric dynamics, auroral current systems, or hybrid scenarios.

\subsection{The detection of the ECM Emission as signature of plasma processes in the magnetosphere of the Ultra Cool Dwarfs}
Many UCDs exhibit coherent radio bursts \citep{Berger2002,Burgasser_Putman2005,hallinan2008,Route2016,Kao2016} generated by the electron cyclotron maser emission (ECME) mechanism.

The coherent emission is highly directive and transient, often appearing at specific rotational phases. Such events are short-lived, from a few minutes to a few hours, as a result of rotational modulation, and are not continuously present. Their detection therefore depends on the fraction of time during which these bursts are active. Despite their transient nature, coherent bursts provide an exceptional diagnostic of the magnetospheric structure and plasma environment in UCDs.

These coherent bursts are typically one to two orders of magnitude brighter than the quiescent component and show strong circular polarization, with flux densities ranging from tens of $\mu$Jy \citep{Lynch2016} to several tens of mJy \citep{Burgasser_Putman2005}, making them easily identifiable in radio observations, in particular using the future high sensitivity measurements that will performed by SKA.

Using the SKAO sensitivity calculator for Band 2, we estimate the detectability of coherent ECME bursts assuming a characteristic burst duration of approximately 10 minutes and a corresponding snapshot sensitivity. Since coherent bursts are transient and strongly time-dependent, detectability depends on the instantaneous sensitivity achieved over the burst duration rather than the total integration time of a survey map.

Under these assumptions, any UCD within approximately 300 pc observed during an active coherent emission phase would lie above the nominal detection threshold.

Assuming a 70\% sky coverage, an average burst duty cycle of approximately 5\%, and neglecting additional geometric beaming effects, we estimate that the SKA could detect of the order of 10,000 coherent radio bursts from UCDs. These estimates should be interpreted as illustrative order-of-magnitude forecasts, since the intrinsic burst occurrence rate, burst duration distribution, and beaming fraction remain uncertain.

The possible detection of intense, highly polarized radio emissions from UCDs at a distance of less than 100~pc will be useful to support the search for the weak radio emission counterpart produced by the incoherent gyro-synchrotron.  

\begin{figure}
    \centering
    \includegraphics[width=1\linewidth]{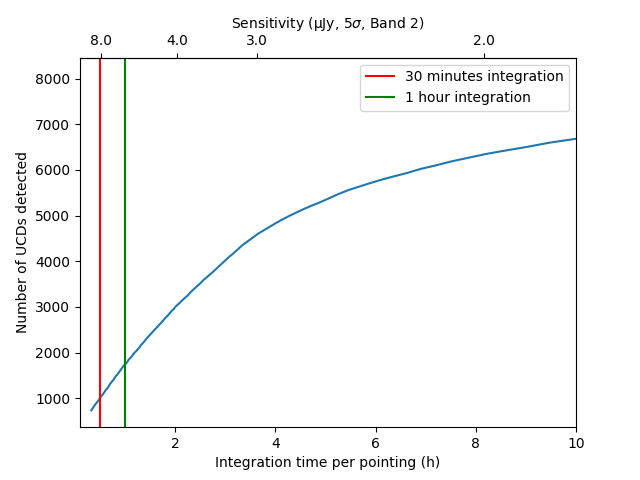}
    \caption{Expected numbers of UCDs detected with a 70\% sky survey in Band 2 carried with AA4 SKA as a function of the integration time per pointing (bottom x-axis) and the $5\sigma$ sensitivity (top x-axis)}
    \label{fig:3Numbers}
\end{figure}

\subsection{SKA's detection capability of incoherent emission from UCDs and Statistical Analysis}
To predict the number of UCDs that could be detected by the SKA through their stable incoherent emission, we will use the space density estimates from \citet{best2024} and \citet{cruz2007}, calculating the total UCD population within a 100 pc radius and the 15\% occurrence rate of incoherent radio emission from \citep{kao2024}. As said in section \ref{subsec:incoherent_ucd}, this gives an expected $\sim$12,500 radio-emitting UCDs in this volume.
For our detectability forecast, we adopt the radio luminosity limits measured by \citep{kao2024}:
$$L_{\mathrm{min}} = 10^{12.7} \ \mathrm{erg \ s^{-1} \ Hz^{-1}}, \quad L_{\mathrm{max}} = 10^{13.6} \ \mathrm{erg \ s^{-1} \ Hz^{-1}}$$

and assume a uniform distribution of luminosities between these bounds. We then compute the expected flux densities by slicing the spherical volume with radius 100~pc into concentric shells and applying
$$S_\nu = \frac{L_\nu}{4\pi d^2}$$
for each luminosity–distance combination, weighted by the shell volume $4\pi d^2 \, \Delta d$.
Fig. \ref{fig:3distri} shows the resulting density distribution of sources as a function of flux density. The distribution peaks at a few $\mathrm{\mu}$Jy, reflecting the prevalence of moderately luminous emitters, and declines smoothly toward higher fluxes corresponding to the upper end of the luminosity range. The adopted luminosity distribution and occurrence fractions are intended as simplified assumptions for illustrative forecasting purposes and do not account for possible population substructures or variability-dependent selection effects.

Using the SKAO sensitivity calculator for Band 2, we estimated the number of UCDs detectable in a wide-area survey as a function of integration time per pointing, adopting the luminosity distribution and spatial model described above. Fig. \ref{fig:3Numbers} presents the results for a 70\% sky survey with the AA4 SKA configuration, showing the detection curve together with reference lines for 30-minute and 1-hour integrations. The top axis indicates the corresponding 5$\sigma$ point-source sensitivities, ranging from $\sim8\,\mathrm{\mu}$Jy for the shortest exposures to $\sim1.5\,\mathrm{\mu}$Jy for the deepest considered. For short integrations, the number of detectable UCDs grows rapidly, exceeding 900 sources for 0.5 h per pointing. The curve flattens toward longer integrations, reaching over 7000 detections at 10 h per pointing. These results demonstrate that even moderate integration times per pointing will allow the SKA surveys to uncover thousands of new incoherent radio-emitting UCDs. Large-area continuum surveys conducted with the SKAO will provide an ideal framework for the systematic discovery of radio-emitting stellar populations, complementing dedicated Galactic Plane surveys at higher radio frequencies \citep{Traficante1.2026.SKA}.

\begin{figure}
    \centering
    \includegraphics[width=1\linewidth]{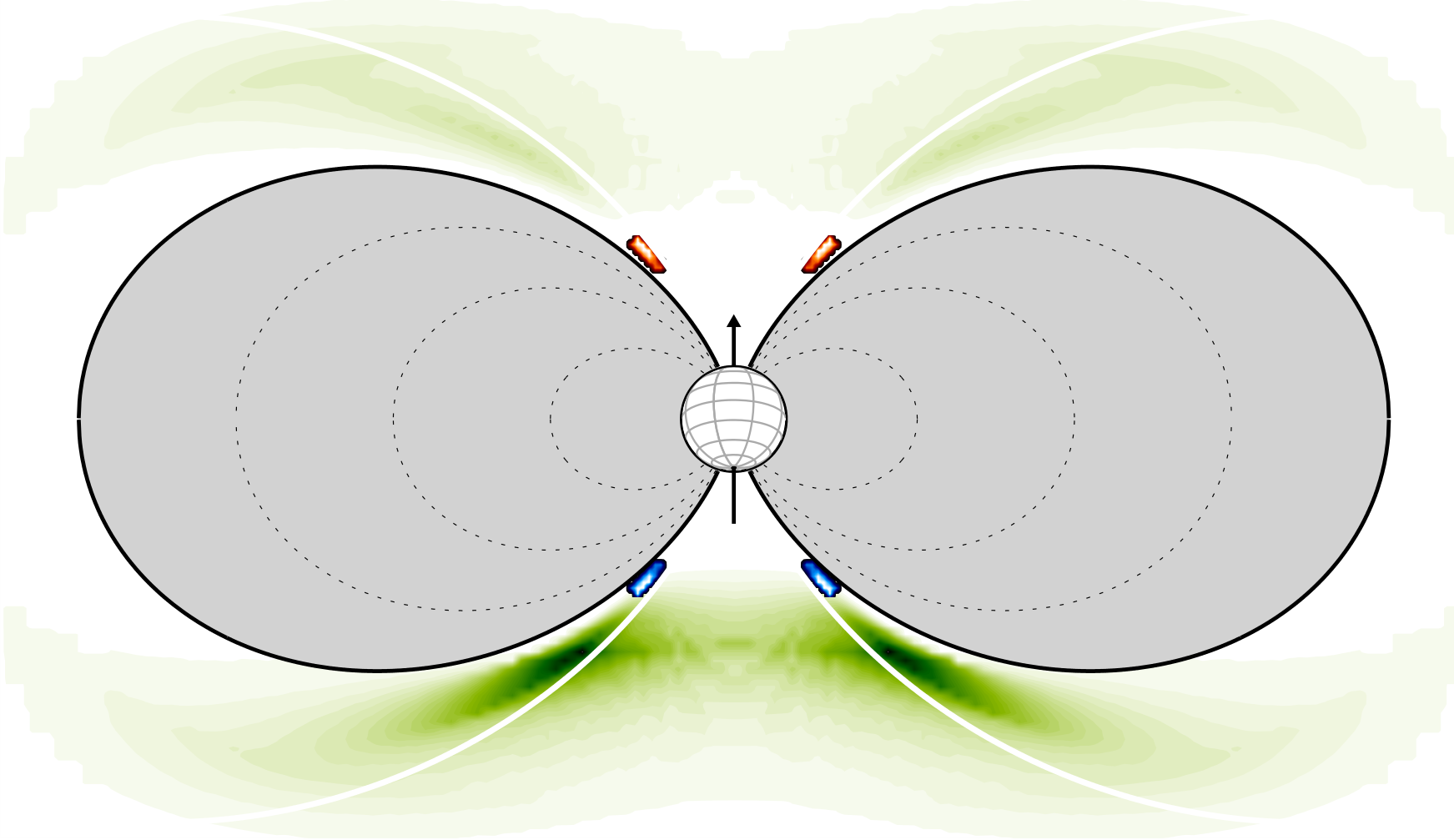}
    \caption{
    Cartoon illustrating the two radio emission components originating within a typical, well-ordered dipolar magnetosphere. The incoherent radio emission, produced by energetic electrons trapped in radiation belts, is shaded in green. The colored spots indicate the sources of the coherent ECME emission arising from the northern (red) and southern (blue) auroral rings. The radiation beam pattern follows the tangent plane beaming model described in Fig.~\ref{tp}. Each ECME source is fully circularly polarized, with the northern and southern auroral emissions exhibiting opposite polarization senses. Figure adapted from \citet{leto2020b}.
}
    \label{fig:scenario}
\end{figure}

\section{SYNERGY}

\subsection{VLBI observations of new nearby UCDs detected by SKA}

The M8.5-type star TVLM~513--46546 (hereafter TVLM~513) was the first UCD detected using very long baseline interferometry (VLBI) techniques \citep{Forbrich2009}. Located at a distance of about 10~pc, its radio magnetosphere was marginally resolved through combined VLBA, VLA, and GBT observations, marking the first direct spatial constraint on a UCD magnetosphere.

The ability of VLBI to directly image the radio magnetospheres of UCDs was later confirmed by \citet{Climent2023} and \citet{Kao2023}. The VLBI maps of the M6 dwarf LSR~J1835+3259, located approximately 6~pc from Earth, clearly reveal the presence of radiation belts producing incoherent emission, together with a central, compact, and highly polarized source corresponding to the region where the ECME mechanism amplifies the coherent radio component.

These spatially resolved measurements of LSR~J1835+3259 provide direct observational confirmation of the scenario illustrated in Fig.~\ref{fig:scenario}, where both the radiation belts and the auroral rings are clearly distinguished. The combination of the wide bandwidth and high sensitivity offered by the future SKA in its AA4 configuration, together with the milliarcsecond-scale spatial resolution achievable with VLBI arrays, will provide a powerful means to study plasma processes within UCD magnetospheres and to search for potential exoplanetary companions through their magnetospheric interactions.

Existing VLBI observations \citep{Forbrich2009,Climent2023,Kao2023} indicate that a distance of roughly 10~pc represents the practical limit for resolving UCD radio magnetospheres. The forthcoming SKA all-sky surveys, with their exceptional sensitivity, are expected to uncover a large number of nearby, radio-loud UCDs. The discovery of such sources will be particularly valuable for planning targeted, high-resolution VLBI observations aimed at resolving their magnetospheres and probing the mechanisms responsible for their non-thermal radio emission. The combination of SKA sensitivity and VLBI resolution will not only resolve magnetospheric structures but may also enable companion searches through precision astrometry \citep{Curiel.1.2026.SKA}.

\section*{SKA outlook for this science case}

The analysis of coherent and incoherent radio emission from stars and substellar objects hosting ordered magnetospheres suggests that related magnetospheric processes may operate across very different stellar mass regimes. In particular, the observed relationship between radio luminosity and centrifugal breakout power in hot magnetic stars, together with analogous trends observed in UCDs, points toward an important role played by large-scale magnetic topology and rapid rotation.

The Square Kilometre Array (SKA) will provide the sensitivity, bandwidth, and survey scale required to test these scenarios under a broad range of physical conditions. Combined with VLBI observations, SKA measurements will enable detailed studies of magnetospheric geometry, plasma distributions, and particle acceleration processes, helping establish a quantitative observational framework for magnetospheric radio emission from stars, brown dwarfs, and star–planet systems.

\bibliography{chapter}

@ARTICLE{trigilio2000,
       author = {{Trigilio}, C. and {Leto}, P. and {Leone}, F. and {Umana}, G. and {Buemi}, C.},
        title = "{Coherent radio emission from the magnetic chemically peculiar star CU Virginis}",
      journal = {\aap},
         year = 2000,
        month = oct,
       volume = {362},
        pages = {281-288},
          doi = {10.48550/arXiv.astro-ph/0007097},
archivePrefix = {arXiv},
       eprint = {astro-ph/0007097},
 primaryClass = {astro-ph},
       adsurl = {https://ui.adsabs.harvard.edu/abs/2000A&A...362..281T}
}

@ARTICLE{kao2023,
       author = {{Kao}, Melodie M. and {Mioduszewski}, Amy J. and {Villadsen}, Jackie and {Shkolnik}, Evgenya L.},
        title = "{Resolved imaging confirms a radiation belt around an ultracool dwarf}",
      journal = {\nat},
         year = 2023,
        month = jul,
       volume = {619},
       number = {7969},
        pages = {272-275},
          doi = {10.1038/s41586-023-06138-w},
archivePrefix = {arXiv},
       eprint = {2302.12841},
 primaryClass = {astro-ph.EP},
       adsurl = {https://ui.adsabs.harvard.edu/abs/2023Natur.619..272K}
}

@ARTICLE{shultz2018,
       author = {{Shultz}, M.~E. and {Wade}, G.~A. and {Rivinius}, Th and {Neiner}, C. and
         {Alecian}, E. and {Bohlender}, D. and {Monin}, D. and {Sikora}, J. and
         {MiMeS Collaboration} and {BinaMIcS Collaboration}},
        title = "{The magnetic early B-type stars I: magnetometry and rotation}",
      journal = {\mnras},
         year = "2018",
        month = "Apr",
       volume = {475},
       number = {4},
        pages = {5144-5178},
          doi = {10.1093/mnras/sty103},
archivePrefix = {arXiv},
       eprint = {1801.02924},
 primaryClass = {astro-ph.SR},
       adsurl = {https://ui.adsabs.harvard.edu/abs/2018MNRAS.475.5144S}
}

@ARTICLE{shultz2019,
       author = {{Shultz}, M.~E. and {Wade}, G.~A. and {Rivinius}, Th and {Alecian}, E. and {Neiner}, C. and {Petit}, V. and {Owocki}, S. and {ud-Doula}, A. and {Kochukhov}, O. and {Bohlender}, D. and {Keszthelyi}, Z. and {MiMeS Collaboration} and {BinaMIcS Collaboration}},
        title = "{The magnetic early B-type stars - III. A main-sequence magnetic, rotational, and magnetospheric biography}",
      journal = {\mnras},
         year = 2019,
        month = nov,
       volume = {490},
       number = {1},
        pages = {274-295},
          doi = {10.1093/mnras/stz2551},
archivePrefix = {arXiv},
       eprint = {1909.02530},
 primaryClass = {astro-ph.SR},
       adsurl = {https://ui.adsabs.harvard.edu/abs/2019MNRAS.490..274S}
}

@ARTICLE{sikora2019a,
       author = {{Sikora}, J. and {Wade}, G.~A. and {Power}, J. and {Neiner}, C.},
        title = "{A volume-limited survey of mCP stars within 100 pc - I. Fundamental parameters and chemical abundances}",
      journal = {\mnras},
         year = 2019,
        month = feb,
       volume = {483},
       number = {2},
        pages = {2300-2324},
          doi = {10.1093/mnras/sty3105},
archivePrefix = {arXiv},
       eprint = {1811.05633},
 primaryClass = {astro-ph.SR},
       adsurl = {https://ui.adsabs.harvard.edu/abs/2019MNRAS.483.2300S}
}

@ARTICLE{townsend2005,
       author = {{Townsend}, R.~H.~D. and {Owocki}, S.~P.},
        title = "{A rigidly rotating magnetosphere model for circumstellar emission from magnetic OB stars}",
      journal = {\mnras},
         year = 2005,
        month = feb,
       volume = {357},
       number = {1},
        pages = {251-264},
          doi = {10.1111/j.1365-2966.2005.08642.x},
archivePrefix = {arXiv},
       eprint = {astro-ph/0408565},
 primaryClass = {astro-ph},
       adsurl = {https://ui.adsabs.harvard.edu/abs/2005MNRAS.357..251T}
}

@ARTICLE{uddoula2006,
       author = {{ud-Doula}, Asif and {Townsend}, Richard H.~D. and {Owocki}, Stanley P.},
        title = "{Centrifugal Breakout of Magnetically Confined Line-driven Stellar Winds}",
      journal = {\apjl},
         year = 2006,
        month = apr,
       volume = {640},
       number = {2},
        pages = {L191-L194},
          doi = {10.1086/503382},
archivePrefix = {arXiv},
       eprint = {astro-ph/0601193},
 primaryClass = {astro-ph},
       adsurl = {https://ui.adsabs.harvard.edu/abs/2006ApJ...640L.191U}
}

@ARTICLE{hoshino2001,
       author = {{Hoshino}, M. and {Mukai}, T. and {Terasawa}, T. and {Shinohara}, I.},
        title = "{Suprathermal electron acceleration in magnetic reconnection}",
      journal = {\jgr},
         year = 2001,
        month = nov,
       volume = {106},
       number = {A11},
        pages = {25979-25998},
          doi = {10.1029/2001JA900052},
       adsurl = {https://ui.adsabs.harvard.edu/abs/2001JGR...10625979H}
}

@ARTICLE{zweibel2009,
       author = {{Zweibel}, Ellen G. and {Yamada}, Masaaki},
        title = "{Magnetic Reconnection in Astrophysical and Laboratory Plasmas}",
      journal = {\araa},
         year = 2009,
        month = sep,
       volume = {47},
       number = {1},
        pages = {291-332},
          doi = {10.1146/annurev-astro-082708-101726},
       adsurl = {https://ui.adsabs.harvard.edu/abs/2009ARA&A..47..291Z}
}

@ARTICLE{trigilio2004,
       author = {{Trigilio}, C. and {Leto}, P. and {Umana}, G. and {Leone}, F. and {Buemi}, C.~S.},
        title = "{A three-dimensional model for the radio emission of magnetic chemically peculiar stars}",
      journal = {\aap},
         year = 2004,
        month = may,
       volume = {418},
        pages = {593-605},
          doi = {10.1051/0004-6361:20040060},
archivePrefix = {arXiv},
       eprint = {astro-ph/0402432},
 primaryClass = {astro-ph},
       adsurl = {https://ui.adsabs.harvard.edu/abs/2004A&A...418..593T}
}

@ARTICLE{leto2006,
       author = {{Leto}, P. and {Trigilio}, C. and {Buemi}, C.~S. and {Umana}, G. and {Leone}, F.},
        title = "{Stellar magnetosphere reconstruction from radio data. Multi-frequency VLA observations and 3D-simulations of <ASTROBJ>CU Virginis</ASTROBJ>}",
      journal = {\aap},
         year = 2006,
        month = nov,
       volume = {458},
       number = {3},
        pages = {831-839},
          doi = {10.1051/0004-6361:20054511},
archivePrefix = {arXiv},
       eprint = {astro-ph/0610395},
 primaryClass = {astro-ph},
       adsurl = {https://ui.adsabs.harvard.edu/abs/2006A&A...458..831L}
}

@ARTICLE{leto2016,
       author = {{Leto}, P. and {Trigilio}, C. and {Buemi}, C.~S. and {Umana}, G. and {Ingallinera}, A. and {Cerrigone}, L.},
        title = "{3D modelling of stellar auroral radio emission}",
      journal = {\mnras},
         year = 2016,
        month = jun,
       volume = {459},
       number = {2},
        pages = {1159-1169},
          doi = {10.1093/mnras/stw639},
archivePrefix = {arXiv},
       eprint = {1603.02423},
 primaryClass = {astro-ph.SR},
       adsurl = {https://ui.adsabs.harvard.edu/abs/2016MNRAS.459.1159L}
}

@ARTICLE{leto2017,
       author = {{Leto}, P. and {Trigilio}, C. and {Oskinova}, L. and {Ignace}, R. and {Buemi}, C.~S. and {Umana}, G. and {Ingallinera}, A. and {Todt}, H. and {Leone}, F.},
        title = "{The detection of variable radio emission from the fast rotating magnetic hot B-star HR 7355 and evidence for its X-ray aurorae}",
      journal = {\mnras},
         year = 2017,
        month = may,
       volume = {467},
       number = {3},
        pages = {2820-2833},
          doi = {10.1093/mnras/stx267},
archivePrefix = {arXiv},
       eprint = {1701.07679},
 primaryClass = {astro-ph.SR},
       adsurl = {https://ui.adsabs.harvard.edu/abs/2017MNRAS.467.2820L}
}

@ARTICLE{owocki2022,
       author = {{Owocki}, S.~P. and {Shultz}, M.~E. and {ud-Doula}, A. and {Chandra}, P. and {Das}, B. and {Leto}, P.},
        title = "{Centrifugal breakout reconnection as the electron acceleration mechanism powering the radio magnetospheres of early-type stars}",
      journal = {\mnras},
         year = 2022,
        month = jun,
       volume = {513},
       number = {1},
        pages = {1449-1458},
          doi = {10.1093/mnras/stac341},
archivePrefix = {arXiv},
       eprint = {2202.05449},
 primaryClass = {astro-ph.SR},
       adsurl = {https://ui.adsabs.harvard.edu/abs/2022MNRAS.513.1449O}
}

@ARTICLE{trigilio2011,
       author = {{Trigilio}, Corrado and {Leto}, Paolo and {Umana}, Grazia and {Buemi}, Carla S. and {Leone}, Francesco},
        title = "{Auroral Radio Emission from Stars: The Case of CU Virginis}",
      journal = {\apjl},
         year = 2011,
        month = sep,
       volume = {739},
       number = {1},
          eid = {L10},
        pages = {L10},
          doi = {10.1088/2041-8205/739/1/L10},
archivePrefix = {arXiv},
       eprint = {1104.3268},
 primaryClass = {astro-ph.SR},
       adsurl = {https://ui.adsabs.harvard.edu/abs/2011ApJ...739L..10T}
}

@ARTICLE{das2020,
       author = {{Das}, Barnali and {Mondal}, Surajit and {Chandra}, Poonam},
        title = "{A 3D Framework to Explore the Propagation Effects in Stars Exhibiting Electron Cyclotron Maser Emission}",
      journal = {\apj},
         year = 2020,
        month = sep,
       volume = {900},
       number = {2},
          eid = {156},
        pages = {156},
          doi = {10.3847/1538-4357/aba8fd},
archivePrefix = {arXiv},
       eprint = {2007.06822},
 primaryClass = {astro-ph.SR},
       adsurl = {https://ui.adsabs.harvard.edu/abs/2020ApJ...900..156D}
}

@ARTICLE{das2021,
       author = {{Das}, Barnali and {Chandra}, Poonam},
        title = "{Ultra-wideband, Multiepoch Radio Study of the First Discovered ``Main-sequence Radio Pulse Emitter'' CU Vir}",
      journal = {\apj},
         year = 2021,
        month = nov,
       volume = {921},
       number = {1},
          eid = {9},
        pages = {9},
          doi = {10.3847/1538-4357/ac1075},
archivePrefix = {arXiv},
       eprint = {2107.00849},
 primaryClass = {astro-ph.SR},
       adsurl = {https://ui.adsabs.harvard.edu/abs/2021ApJ...921....9D}
}

@ARTICLE{climent2023,
       author = {{Climent}, J.~B. and {Guirado}, J.~C. and {P{\'e}rez-Torres}, M. and {Marcaide}, J.~M. and {Pe{\~n}a-Mo{\~n}ino}, L.},
        title = "{Evidence for a radiation belt around a brown dwarf}",
      journal = {Science},
         year = 2023,
        month = sep,
       volume = {381},
       number = {6662},
        pages = {1120-1124},
          doi = {10.1126/science.adg6635},
archivePrefix = {arXiv},
       eprint = {2303.06453},
 primaryClass = {astro-ph.SR},
       adsurl = {https://ui.adsabs.harvard.edu/abs/2023Sci...381.1120C}
}

@ARTICLE{hallinan2008,
       author = {{Hallinan}, G. and {Antonova}, A. and {Doyle}, J.~G. and {Bourke}, S. and {Lane}, C. and {Golden}, A.},
        title = "{Confirmation of the Electron Cyclotron Maser Instability as the Dominant Source of Radio Emission from Very Low Mass Stars and Brown Dwarfs}",
      journal = {\apj},
         year = 2008,
        month = sep,
       volume = {684},
       number = {1},
        pages = {644-653},
          doi = {10.1086/590360},
archivePrefix = {arXiv},
       eprint = {0805.4010},
 primaryClass = {astro-ph},
       adsurl = {https://ui.adsabs.harvard.edu/abs/2008ApJ...684..644H}
}

@ARTICLE{treumann2006,
       author = {{Treumann}, Rudolf A.},
        title = "{The electron-cyclotron maser for astrophysical application}",
      journal = {\aapr},
         year = 2006,
        month = aug,
       volume = {13},
       number = {4},
        pages = {229-315},
          doi = {10.1007/s00159-006-0001-y},
       adsurl = {https://ui.adsabs.harvard.edu/abs/2006A&ARv..13..229T}
}

@ARTICLE{williams2014,
       author = {{Williams}, P.~K.~G. and {Cook}, B.~A. and {Berger}, E.},
        title = "{Trends in Ultracool Dwarf Magnetism. I. X-Ray Suppression and Radio Enhancement}",
      journal = {\apj},
         year = 2014,
        month = apr,
       volume = {785},
       number = {1},
          eid = {9},
        pages = {9},
          doi = {10.1088/0004-637X/785/1/9},
archivePrefix = {arXiv},
       eprint = {1310.6757},
 primaryClass = {astro-ph.SR},
       adsurl = {https://ui.adsabs.harvard.edu/abs/2014ApJ...785....9W}
}

@ARTICLE{reiners2007,
       author = {{Reiners}, A. and {Basri}, G.},
        title = "{The First Direct Measurements of Surface Magnetic Fields on Very Low Mass Stars}",
      journal = {\apj},
         year = 2007,
        month = feb,
       volume = {656},
       number = {2},
        pages = {1121-1135},
          doi = {10.1086/510304},
archivePrefix = {arXiv},
       eprint = {astro-ph/0610365},
 primaryClass = {astro-ph},
       adsurl = {https://ui.adsabs.harvard.edu/abs/2007ApJ...656.1121R}
}

@ARTICLE{metodieva2017,
       author = {{Metodieva}, Y.~T. and {Kuznetsov}, A.~A. and {Antonova}, A.~E. and {Doyle}, J.~G. and {Ramsay}, G. and {Wu}, K.},
        title = "{Modelling the environment around five ultracool dwarfs via the radio domain}",
      journal = {\mnras},
         year = 2017,
        month = feb,
       volume = {465},
       number = {2},
        pages = {1995-2009},
          doi = {10.1093/mnras/stw2597},
archivePrefix = {arXiv},
       eprint = {1610.02989},
 primaryClass = {astro-ph.SR},
       adsurl = {https://ui.adsabs.harvard.edu/abs/2017MNRAS.465.1995M}
}

@ARTICLE{mclean2012,
       author = {{McLean}, M. and {Berger}, E. and {Reiners}, A.},
        title = "{The Radio Activity-Rotation Relation of Ultracool Dwarfs}",
      journal = {\apj},
         year = 2012,
        month = feb,
       volume = {746},
       number = {1},
          eid = {23},
        pages = {23},
          doi = {10.1088/0004-637X/746/1/23},
archivePrefix = {arXiv},
       eprint = {1108.0415},
 primaryClass = {astro-ph.SR},
       adsurl = {https://ui.adsabs.harvard.edu/abs/2012ApJ...746...23M}
}

@ARTICLE{cruz2007,
       author = {{Cruz}, Kelle L. and {Reid}, I. Neill and {Kirkpatrick}, J. Davy and {Burgasser}, Adam J. and {Liebert}, James and {Solomon}, Adam R. and {Schmidt}, Sarah J. and {Allen}, Peter R. and {Hawley}, Suzanne L. and {Covey}, Kevin R.},
        title = "{Meeting the Cool Neighbors. IX. The Luminosity Function of M7-L8 Ultracool Dwarfs in the Field}",
      journal = {\aj},
         year = 2007,
        month = feb,
       volume = {133},
       number = {2},
        pages = {439-467},
          doi = {10.1086/510132},
archivePrefix = {arXiv},
       eprint = {astro-ph/0609648},
 primaryClass = {astro-ph},
       adsurl = {https://ui.adsabs.harvard.edu/abs/2007AJ....133..439C}
}

@ARTICLE{Best2024,
       author = {{Best}, William M.~J. and {Sanghi}, Aniket and {Liu}, Michael C. and {Magnier}, Eugene A. and {Dupuy}, Trent J.},
        title = "{A Volume-limited Sample of Ultracool Dwarfs. II. The Substellar Age and Mass Functions in the Solar Neighborhood}",
      journal = {\apj},
         year = 2024,
        month = jun,
       volume = {967},
       number = {2},
          eid = {115},
        pages = {115},
          doi = {10.3847/1538-4357/ad39ef},
archivePrefix = {arXiv},
       eprint = {2401.09535},
 primaryClass = {astro-ph.SR},
       adsurl = {https://ui.adsabs.harvard.edu/abs/2024ApJ...967..115B}
}

@ARTICLE{kao2024,
       author = {{Kao}, Melodie M. and {Shkolnik}, Evgenya L.},
        title = "{The occurrence rate of quiescent radio emission for ultracool dwarfs using a generalized semi-analytical Bayesian framework}",
      journal = {\mnras},
         year = 2024,
        month = jan,
       volume = {527},
       number = {3},
        pages = {6835-6866},
          doi = {10.1093/mnras/stad2272},
archivePrefix = {arXiv},
       eprint = {2306.16460},
 primaryClass = {astro-ph.SR},
       adsurl = {https://ui.adsabs.harvard.edu/abs/2024MNRAS.527.6835K}
}

@ARTICLE{chandra2015,
       author = {{Chandra}, P. and {Wade}, G.~A. and {Sundqvist}, J.~O. and {Oberoi}, D. and
         {Grunhut}, J.~H. and {ud-Doula}, A. and {Petit}, V. and {Cohen}, D.~H. and
         {Oksala}, M.~E. and {David-Uraz}, A.},
        title = "{Detection of 610-MHz radio emission from hot magnetic stars}",
      journal = {\mnras},
         year = "2015",
        month = "Sep",
       volume = {452},
       number = {2},
        pages = {1245-1253},
          doi = {10.1093/mnras/stv1378},
archivePrefix = {arXiv},
       eprint = {1505.02139},
 primaryClass = {astro-ph.SR},
       adsurl = {https://ui.adsabs.harvard.edu/abs/2015MNRAS.452.1245C}
}

@ARTICLE{lenc2018,
       author = {{Lenc}, Emil and {Murphy}, Tara and {Lynch}, C.~R. and {Kaplan}, D.~L. and
         {Zhang}, S.~N.},
        title = "{An all-sky survey of circular polarization at 200 MHz}",
      journal = {\mnras},
         year = "2018",
        month = "Aug",
       volume = {478},
       number = {2},
        pages = {2835-2849},
          doi = {10.1093/mnras/sty1304},
archivePrefix = {arXiv},
       eprint = {1805.05482},
 primaryClass = {astro-ph.GA},
       adsurl = {https://ui.adsabs.harvard.edu/abs/2018MNRAS.478.2835L}
}

@ARTICLE{das2018,
       author = {{Das}, Barnali and {Chandra}, Poonam and {Wade}, Gregg A.},
        title = "{Discovery of electron cyclotron MASER emission from the magnetic Bp star HD 133880 with the Giant Metrewave Radio Telescope}",
      journal = {\mnras},
         year = "2018",
        month = "Feb",
       volume = {474},
       number = {1},
        pages = {L61-L65},
          doi = {10.1093/mnrasl/slx193},
archivePrefix = {arXiv},
       eprint = {1711.09836},
 primaryClass = {astro-ph.SR},
       adsurl = {https://ui.adsabs.harvard.edu/abs/2018MNRAS.474L..61D}
}

@ARTICLE{leto2019,
       author = {{Leto}, P. and {Trigilio}, C. and {Oskinova}, L.~M. and {Ignace}, R. and
         {Buemi}, C.~S. and {Umana}, G. and {Cavallaro}, F. and
         {Ingallinera}, A. and {Bufano}, F. and {Phillips}, N.~M. and
         {Agliozzo}, C. and {Cerrigone}, L. and {Todt}, H. and {Riggi}, S. and
         {Leone}, F.},
        title = "{The polarization mode of the auroral radio emission from the early-type star HD 142301}",
      journal = {\mnras},
         year = "2019",
        month = "Jan",
       volume = {482},
       number = {1},
        pages = {L4-L8},
          doi = {10.1093/mnrasl/sly179},
archivePrefix = {arXiv},
       eprint = {1809.07504},
 primaryClass = {astro-ph.SR},
       adsurl = {https://ui.adsabs.harvard.edu/abs/2019MNRAS.482L...4L}
}

@ARTICLE{das2019a,
       author = {{Das}, Barnali and {Chandra}, Poonam and {Shultz}, Matt E. and
         {Wade}, Gregg A.},
        title = "{Detection of Coherent Emission from the Bp Star HD 142990 at uGMRT Frequencies}",
      journal = {\apj},
         year = "2019",
        month = "Jun",
       volume = {877},
       number = {2},
          eid = {123},
        pages = {123},
          doi = {10.3847/1538-4357/ab1b12},
archivePrefix = {arXiv},
       eprint = {1904.08359},
 primaryClass = {astro-ph.SR},
       adsurl = {https://ui.adsabs.harvard.edu/abs/2019ApJ...877..123D}
}

@ARTICLE{das2019b,
       author = {{Das}, Barnali and {Chandra}, Poonam and {Shultz}, Matt E. and {Wade}, Gregg A.},
        title = "{The fifth main-sequence magnetic B-type star showing coherent radio emission: Is this really a rare phenomenon?}",
      journal = {\mnras},
         year = 2019,
        month = oct,
       volume = {489},
       number = {1},
        pages = {L102-L107},
          doi = {10.1093/mnrasl/slz137},
archivePrefix = {arXiv},
       eprint = {1908.09110},
 primaryClass = {astro-ph.SR},
       adsurl = {https://ui.adsabs.harvard.edu/abs/2019MNRAS.489L.102D}
}

@ARTICLE{leto2020a,
       author = {{Leto}, P. and {Trigilio}, C. and {Leone}, F. and {Pillitteri}, I. and {Buemi}, C.~S. and {Fossati}, L. and {Cavallaro}, F. and {Oskinova}, L.~M. and {Ignace}, R. and {Krti{\v{c}}ka}, J. and {Umana}, G. and {Catanzaro}, G. and {Ingallinera}, A. and {Bufano}, F. and {Agliozzo}, C. and {Phillips}, N.~M. and {Cerrigone}, L. and {Riggi}, S. and {Loru}, S. and {Munari}, M. and {Gangi}, M. and {Giarrusso}, M. and {Robrade}, J.},
        title = "{Evidence for radio and X-ray auroral emissions from the magnetic B-type star {\ensuremath{\rho}} Oph A}",
      journal = {\mnras},
         year = 2020,
        month = apr,
       volume = {493},
       number = {4},
        pages = {4657-4676},
          doi = {10.1093/mnras/staa587},
archivePrefix = {arXiv},
       eprint = {2002.09251},
 primaryClass = {astro-ph.SR},
       adsurl = {https://ui.adsabs.harvard.edu/abs/2020MNRAS.493.4657L}
}

@ARTICLE{leto2020b,
       author = {{Leto}, P. and {Trigilio}, C. and {Buemi}, C.~S. and {Leone}, F. and {Pillitteri}, I. and {Fossati}, L. and {Cavallaro}, F. and {Oskinova}, L.~M. and {Ignace}, R. and {Krti{\v{c}}ka}, J. and {Umana}, G. and {Catanzaro}, G. and {Ingallinera}, A. and {Bufano}, F. and {Riggi}, S. and {Cerrigone}, L. and {Loru}, S. and {Schillir{\'o}}, F. and {Agliozzo}, C. and {Phillips}, N.~M. and {Giarrusso}, M. and {Robrade}, J.},
        title = "{The auroral radio emission of the magnetic B-type star {\ensuremath{\rho}}-=OphC}",
      journal = {\mnras},
         year = 2020,
        month = sep,
       volume = {499},
       number = {1},
        pages = {L72-L76},
          doi = {10.1093/mnrasl/slaa157},
archivePrefix = {arXiv},
       eprint = {2009.02363},
 primaryClass = {astro-ph.SR},
       adsurl = {https://ui.adsabs.harvard.edu/abs/2020MNRAS.499L..72L}
}

@ARTICLE{pritchard2021,
       author = {{Pritchard}, Joshua and {Murphy}, Tara and {Zic}, Andrew and {Lynch}, Christene and {Heald}, George and {Kaplan}, David L. and {Anderson}, Craig and {Banfield}, Julie and {Hale}, Catherine and {Hotan}, Aidan and {Lenc}, Emil and {Leung}, James K. and {McConnell}, David and {Moss}, Vanessa A. and {Raja}, Wasim and {Stewart}, Adam J. and {Whiting}, Matthew},
        title = "{A circular polarization survey for radio stars with the Australian SKA Pathfinder}",
      journal = {\mnras},
         year = 2021,
        month = apr,
       volume = {502},
       number = {4},
        pages = {5438-5454},
          doi = {10.1093/mnras/stab299},
archivePrefix = {arXiv},
       eprint = {2102.01801},
 primaryClass = {astro-ph.SR},
       adsurl = {https://ui.adsabs.harvard.edu/abs/2021MNRAS.502.5438P}
}

@ARTICLE{das2022a,
       author = {{Das}, Barnali and {Chandra}, Poonam and {Shultz}, Matt E. and {Wade}, Gregg A. and {Sikora}, James and {Kochukhov}, Oleg and {Neiner}, Coralie and {Oksala}, Mary E. and {Alecian}, Evelyne},
        title = "{Discovery of Eight ``Main-sequence Radio Pulse Emitters'' Using the GMRT: Clues to the Onset of Coherent Radio Emission in Hot Magnetic Stars}",
      journal = {\apj},
         year = 2022,
        month = feb,
       volume = {925},
       number = {2},
          eid = {125},
        pages = {125},
          doi = {10.3847/1538-4357/ac2576},
archivePrefix = {arXiv},
       eprint = {2109.04043},
 primaryClass = {astro-ph.SR},
       adsurl = {https://ui.adsabs.harvard.edu/abs/2022ApJ...925..125D}
}

@ARTICLE{das2022b,
       author = {{Das}, Barnali and {Chandra}, Poonam and {Petit}, V{\'e}ronique},
        title = "{What leads to premature upper cut-off frequencies of auroral radio emission from hot magnetic stars?}",
      journal = {\mnras},
         year = 2022,
        month = sep,
       volume = {515},
       number = {2},
        pages = {2008-2014},
          doi = {10.1093/mnras/stac1894},
archivePrefix = {arXiv},
       eprint = {2207.00470},
 primaryClass = {astro-ph.SR},
       adsurl = {https://ui.adsabs.harvard.edu/abs/2022MNRAS.515.2008D}
}

@ARTICLE{das2022c,
       author = {{Das}, Barnali and {Chandra}, Poonam and {Shultz}, Matt E. and {Leto}, Paolo and {Mikul{\'a}{\v{s}}ek}, Zden{\v{e}}k and {Petit}, V{\'e}ronique and {Wade}, Gregg A.},
        title = "{Testing a scaling relation between coherent radio emission and physical parameters of hot magnetic stars}",
      journal = {\mnras},
         year = 2022,
        month = dec,
       volume = {517},
       number = {4},
        pages = {5756-5769},
          doi = {10.1093/mnras/stac3123},
archivePrefix = {arXiv},
       eprint = {2210.14746},
 primaryClass = {astro-ph.SR},
       adsurl = {https://ui.adsabs.harvard.edu/abs/2022MNRAS.517.5756D}
}

@ARTICLE{biswas2025,
       author = {{Biswas}, Ayan and {Das}, Barnali and {Barron}, James A. and {Wade}, Gregg A. and {Holgado}, Gonzalo},
        title = "{A Nonstop Aurora? The Intriguing Radio Emission from the Rapidly Rotating Magnetic Massive Star HR 5907}",
      journal = {\apj},
         year = 2025,
        month = feb,
       volume = {980},
       number = {2},
          eid = {260},
        pages = {260},
          doi = {10.3847/1538-4357/adae02},
archivePrefix = {arXiv},
       eprint = {2501.10813},
 primaryClass = {astro-ph.SR},
       adsurl = {https://ui.adsabs.harvard.edu/abs/2025ApJ...980..260B}
}

@ARTICLE{zarka2004,
       author = {{Zarka}, P.},
        title = "{Radio and plasma waves at the outer planets}",
      journal = {Advances in Space Research},
         year = 2004,
        month = jan,
       volume = {33},
       number = {11},
        pages = {2045-2060},
          doi = {10.1016/j.asr.2003.07.055},
       adsurl = {https://ui.adsabs.harvard.edu/abs/2004AdSpR..33.2045Z}
}

@ARTICLE{das2025a,
       author = {{Das}, Barnali and {Chandra}, Poonam and {Cotton}, William and {Petit}, V{\'e}ronique},
        title = "{Discoveries of fine structures and secondary pulses in coherent radio emission from a magnetic massive star}",
      journal = {arXiv e-prints},
         year = 2025,
        month = jul,
          eid = {arXiv:2507.03882},
        pages = {arXiv:2507.03882},
          doi = {10.48550/arXiv.2507.03882},
archivePrefix = {arXiv},
       eprint = {2507.03882},
 primaryClass = {astro-ph.SR},
       adsurl = {https://ui.adsabs.harvard.edu/abs/2025arXiv250703882D}
}

@ARTICLE{das2024,
       author = {{Das}, Barnali and {Chandra}, Poonam and {Petit}, V{\'e}ronique},
        title = "{Coherent Radio Emission from ``Main-sequence Radio Pulse Emitters'': A New Stellar Diagnostic to Probe 3D Magnetospheric Structures}",
      journal = {\apj},
         year = 2024,
        month = oct,
       volume = {974},
       number = {2},
          eid = {267},
        pages = {267},
          doi = {10.3847/1538-4357/ad71c5},
archivePrefix = {arXiv},
       eprint = {2408.11242},
 primaryClass = {astro-ph.SR},
       adsurl = {https://ui.adsabs.harvard.edu/abs/2024ApJ...974..267D}
}

@ARTICLE{leto2021,
       author = {{Leto}, P. and {Trigilio}, C. and {Krti{\v{c}}ka}, J. and {Fossati}, L. and {Ignace}, R. and {Shultz}, M.~E. and {Buemi}, C.~S. and {Cerrigone}, L. and {Umana}, G. and {Ingallinera}, A. and {Bordiu}, C. and {Pillitteri}, I. and {Bufano}, F. and {Oskinova}, L.~M. and {Agliozzo}, C. and {Cavallaro}, F. and {Riggi}, S. and {Loru}, S. and {Todt}, H. and {Giarrusso}, M. and {Phillips}, N.~M. and {Robrade}, J. and {Leone}, F.},
        title = "{A scaling relationship for non-thermal radio emission from ordered magnetospheres: from the top of the main sequence to planets}",
      journal = {\mnras},
         year = 2021,
        month = oct,
       volume = {507},
       number = {2},
        pages = {1979-1998},
          doi = {10.1093/mnras/stab2168},
archivePrefix = {arXiv},
       eprint = {2107.11995},
 primaryClass = {astro-ph.SR},
       adsurl = {https://ui.adsabs.harvard.edu/abs/2021MNRAS.507.1979L}
}

@ARTICLE{shultz2022,
       author = {{Shultz}, M.~E. and {Owocki}, S.~P. and {ud-Doula}, A. and {Biswas}, A. and {Bohlender}, D. and {Chandra}, P. and {Das}, B. and {David-Uraz}, A. and {Khalack}, V. and {Kochukhov}, O. and {Landstreet}, J.~D. and {Leto}, P. and {Monin}, D. and {Neiner}, C. and {Rivinius}, Th and {Wade}, G.~A.},
        title = "{MOBSTER - VI. The crucial influence of rotation on the radio magnetospheres of hot stars}",
      journal = {\mnras},
         year = 2022,
        month = apr,
          doi = {10.1093/mnras/stac136},
archivePrefix = {arXiv},
       eprint = {2201.05512},
 primaryClass = {astro-ph.SR},
       adsurl = {https://ui.adsabs.harvard.edu/abs/2022MNRAS.tmp.1099S}
}

@ARTICLE{depater2003,
       author = {{de Pater}, Imke and {Dunn}, David E.},
        title = "{VLA observations of Jupiter's synchrotron radiation at 15 and 22 GHz}",
      journal = {\icarus},
         year = 2003,
        month = jun,
       volume = {163},
       number = {2},
        pages = {449-455},
          doi = {10.1016/S0019-1035(03)00068-X},
       adsurl = {https://ui.adsabs.harvard.edu/abs/2003Icar..163..449D}
}

@ARTICLE{Melrose1982,
       author = {{Melrose}, D.~B. and {Dulk}, G.~A.},
        title = "{Electron-cyclotron masers as the source of certain solar and stellar radio bursts.}",
      journal = {\apj},
         year = 1982,
        month = aug,
       volume = {259},
        pages = {844-858},
          doi = {10.1086/160219},
       adsurl = {https://ui.adsabs.harvard.edu/abs/1982ApJ...259..844M}
}

@ARTICLE{Zarka1998,
       author = {{Zarka}, Philippe},
        title = "{Auroral radio emissions at the outer planets: Observations and theories}",
      journal = {\jgr},
         year = 1998,
        month = sep,
       volume = {103},
       number = {E9},
        pages = {20159-20194},
          doi = {10.1029/98JE01323},
       adsurl = {https://ui.adsabs.harvard.edu/abs/1998JGR...10320159Z}
}

@ARTICLE{Zarka2007pss,
       author = {{Zarka}, Philippe},
        title = "{Plasma interactions of exoplanets with their parent star and associated radio emissions}",
      journal = {\planss},
         year = 2007,
        month = apr,
       volume = {55},
       number = {5},
        pages = {598-617},
          doi = {10.1016/j.pss.2006.05.045},
       adsurl = {https://ui.adsabs.harvard.edu/abs/2007P&SS...55..598Z}
}

@ARTICLE{Shultz2019MNRAS485,
       author = {{Shultz}, M.~E. and {Wade}, G.~A. and {Rivinius}, Th and {Alecian}, E. and {Neiner}, C. and {Petit}, V. and {Wisniewski}, J.~P. and {MiMeS Collaboration} and {BinaMIcS Collaboration}},
        title = "{The magnetic early B-type Stars II: stellar atmospheric parameters in the era of Gaia}",
      journal = {\mnras},
         year = 2019,
        month = may,
       volume = {485},
       number = {2},
        pages = {1508-1527},
          doi = {10.1093/mnras/stz416},
archivePrefix = {arXiv},
       eprint = {1902.02713},
 primaryClass = {astro-ph.SR},
       adsurl = {https://ui.adsabs.harvard.edu/abs/2019MNRAS.485.1508S}
}

@ARTICLE{Renson2009,
       author = {{Renson}, P. and {Manfroid}, J.},
        title = "{Catalogue of Ap, HgMn and Am stars}",
      journal = {\aap},
         year = 2009,
        month = may,
       volume = {498},
       number = {3},
        pages = {961-966},
          doi = {10.1051/0004-6361/200810788},
       adsurl = {https://ui.adsabs.harvard.edu/abs/2009A&A...498..961R}
}

@ARTICLE{Matthews2025,
       author = {{Matthews}, Lynn D.},
        title = "{Radio Stars in the Era of New Observatories}",
      journal = {arXiv e-prints},
         year = 2025,
        month = sep,
          eid = {arXiv:2509.21467},
        pages = {arXiv:2509.21467},
          doi = {10.48550/arXiv.2509.21467},
archivePrefix = {arXiv},
       eprint = {2509.21467},
 primaryClass = {astro-ph.SR},
       adsurl = {https://ui.adsabs.harvard.edu/abs/2025arXiv250921467M}
}

@ARTICLE{Dulk1985,
       author = {{Dulk}, G.~A.},
        title = "{Radio emission from the sun and stars.}",
      journal = {\araa},
         year = 1985,
        month = jan,
       volume = {23},
        pages = {169-224},
          doi = {10.1146/annurev.aa.23.090185.001125},
       adsurl = {https://ui.adsabs.harvard.edu/abs/1985ARA&A..23..169D}
}

@ARTICLE{Leto2012,
       author = {{Leto}, P. and {Trigilio}, C. and {Buemi}, C.~S. and {Leone}, F. and {Umana}, G.},
        title = "{Searching for a CU Virginis-type cyclotron maser from {\ensuremath{\sigma}} Orionis E: the role of the magnetic quadrupole component}",
      journal = {\mnras},
         year = 2012,
        month = jun,
       volume = {423},
       number = {2},
        pages = {1766-1774},
          doi = {10.1111/j.1365-2966.2012.20997.x},
archivePrefix = {arXiv},
       eprint = {1203.6475},
 primaryClass = {astro-ph.SR},
       adsurl = {https://ui.adsabs.harvard.edu/abs/2012MNRAS.423.1766L}
}

@ARTICLE{Badman2015,
       author = {{Badman}, Sarah V. and {Branduardi-Raymont}, Graziella and {Galand}, Marina and {Hess}, S{\'e}bastien L.~G. and {Krupp}, Norbert and {Lamy}, Laurent and {Melin}, Henrik and {Tao}, Chihiro},
        title = "{Auroral Processes at the Giant Planets: Energy Deposition, Emission Mechanisms, Morphology and Spectra}",
      journal = {\ssr},
         year = 2015,
        month = apr,
       volume = {187},
       number = {1-4},
        pages = {99-179},
          doi = {10.1007/s11214-014-0042-x},
       adsurl = {https://ui.adsabs.harvard.edu/abs/2015SSRv..187...99B}
}

@ARTICLE{Babcock1949,
       author = {{Babcock}, H.~W.},
        title = "{Stellar magnetic fields and rotation}",
      journal = {The Observatory},
         year = 1949,
        month = oct,
       volume = {69},
        pages = {191-192},
       adsurl = {https://ui.adsabs.harvard.edu/abs/1949Obs....69..191B}
}

@ARTICLE{Stibbs1950,
       author = {{Stibbs}, D.~W.~N.},
        title = "{A study of the spectrum and magnetic variable star HD 125248}",
      journal = {\mnras},
         year = 1950,
        month = jan,
       volume = {110},
        pages = {395},
          doi = {10.1093/mnras/110.4.395},
       adsurl = {https://ui.adsabs.harvard.edu/abs/1950MNRAS.110..395S}
}

@INPROCEEDINGS{Grunhut2012,
       author = {{Grunhut}, J.~H. and {Wade}, G.~A. and {MiMeS Collaboration}},
        title = "{The Incidence of Magnetic Fields in Massive Stars: An Overview of the MiMeS Survey Component}",
    booktitle = {Proceedings of a Scientific Meeting in Honor of Anthony F. J. Moffat},
         year = 2012,
       editor = {{Drissen}, L. and {Robert}, C. and {St-Louis}, N. and {Moffat}, A.~F.~J.},
       series = {Astronomical Society of the Pacific Conference Series},
       volume = {465},
        month = dec,
        pages = {42},
       adsurl = {https://ui.adsabs.harvard.edu/abs/2012ASPC..465...42G}
}

@ARTICLE{Wade2016,
       author = {{Wade}, G.~A. and {Neiner}, C. and {Alecian}, E. and {Grunhut}, J.~H. and {Petit}, V. and {Batz}, B. de and {Bohlender}, D.~A. and {Cohen}, D.~H. and {Henrichs}, H.~F. and {Kochukhov}, O. and {Landstreet}, J.~D. and {Manset}, N. and {Martins}, F. and {Mathis}, S. and {Oksala}, M.~E. and {Owocki}, S.~P. and {Rivinius}, Th. and {Shultz}, M.~E. and {Sundqvist}, J.~O. and {Townsend}, R.~H.~D. and {ud-Doula}, A. and {Bouret}, J.-C. and {Braithwaite}, J. and {Briquet}, M. and {Carciofi}, A.~C. and {David-Uraz}, A. and {Folsom}, C.~P. and {Fullerton}, A.~W. and {Leroy}, B. and {Marcolino}, W.~L.~F. and {Moffat}, A.~F.~J. and {Naz{\'e}}, Y. and {Louis}, N. St and {Auri{\`e}re}, M. and {Bagnulo}, S. and {Bailey}, J.~D. and {Barb{\'a}}, R.~H. and {Blaz{\`e}re}, A. and {B{\"o}hm}, T. and {Catala}, C. and {Donati}, J.-F. and {Ferrario}, L. and {Harrington}, D. and {Howarth}, I.~D. and {Ignace}, R. and {Kaper}, L. and {L{\"u}ftinger}, T. and {Prinja}, R. and {Vink}, J.~S. and {Weiss}, W.~W. and {Yakunin}, I.},
        title = "{The MiMeS survey of magnetism in massive stars: introduction and overview}",
      journal = {\mnras},
         year = 2016,
        month = feb,
       volume = {456},
       number = {1},
        pages = {2-22},
          doi = {10.1093/mnras/stv2568},
archivePrefix = {arXiv},
       eprint = {1511.08425},
 primaryClass = {astro-ph.SR},
       adsurl = {https://ui.adsabs.harvard.edu/abs/2016MNRAS.456....2W}
}

@ARTICLE{Yadav2015,
       author = {{Yadav}, Rakesh K. and {Christensen}, Ulrich R. and {Morin}, Julien and {Gastine}, Thomas and {Reiners}, Ansgar and {Poppenhaeger}, Katja and {Wolk}, Scott J.},
        title = "{Explaining the Coexistence of Large-scale and Small-scale Magnetic Fields in Fully Convective Stars}",
      journal = {\apjl},
         year = 2015,
        month = nov,
       volume = {813},
       number = {2},
          eid = {L31},
        pages = {L31},
          doi = {10.1088/2041-8205/813/2/L31},
archivePrefix = {arXiv},
       eprint = {1510.05541},
 primaryClass = {astro-ph.SR},
       adsurl = {https://ui.adsabs.harvard.edu/abs/2015ApJ...813L..31Y}
}

@ARTICLE{Donati2006,
       author = {{Donati}, Jean-Fran{\c{c}}ois and {Forveille}, Thierry and {Collier Cameron}, Andrew and {Barnes}, John R. and {Delfosse}, Xavier and {Jardine}, Moira M. and {Valenti}, Jeff A.},
        title = "{The Large-Scale Axisymmetric Magnetic Topology of a Very-Low-Mass Fully Convective Star}",
      journal = {Science},
         year = 2006,
        month = feb,
       volume = {311},
       number = {5761},
        pages = {633-635},
          doi = {10.1126/science.1121102},
archivePrefix = {arXiv},
       eprint = {astro-ph/0602069},
 primaryClass = {astro-ph},
       adsurl = {https://ui.adsabs.harvard.edu/abs/2006Sci...311..633D}
}

@ARTICLE{Morin2010,
       author = {{Morin}, J. and {Donati}, J.-F. and {Petit}, P. and {Delfosse}, X. and {Forveille}, T. and {Jardine}, M.~M.},
        title = "{Large-scale magnetic topologies of late M dwarfs*}",
      journal = {\mnras},
         year = 2010,
        month = oct,
       volume = {407},
       number = {4},
        pages = {2269-2286},
          doi = {10.1111/j.1365-2966.2010.17101.x},
archivePrefix = {arXiv},
       eprint = {1005.5552},
 primaryClass = {astro-ph.SR},
       adsurl = {https://ui.adsabs.harvard.edu/abs/2010MNRAS.407.2269M}
}

@ARTICLE{Route2016,
       author = {{Route}, Matthew and {Wolszczan}, Alexander},
        title = "{The Second Arecibo Search for 5 GHz Radio Flares from Ultracool Dwarfs}",
      journal = {\apj},
         year = 2016,
        month = oct,
       volume = {830},
       number = {2},
          eid = {85},
        pages = {85},
          doi = {10.3847/0004-637X/830/2/85},
archivePrefix = {arXiv},
       eprint = {1608.02480},
 primaryClass = {astro-ph.SR},
       adsurl = {https://ui.adsabs.harvard.edu/abs/2016ApJ...830...85R}
}

@ARTICLE{Lynch2016,
       author = {{Lynch}, C. and {Murphy}, T. and {Ravi}, V. and {Hobbs}, G. and {Lo}, K. and {Ward}, C.},
        title = "{Radio detections of southern ultracool dwarfs}",
      journal = {\mnras},
         year = 2016,
        month = apr,
       volume = {457},
       number = {2},
        pages = {1224-1232},
          doi = {10.1093/mnras/stw050},
archivePrefix = {arXiv},
       eprint = {1601.01749},
 primaryClass = {astro-ph.SR},
       adsurl = {https://ui.adsabs.harvard.edu/abs/2016MNRAS.457.1224L}
}

@ARTICLE{Berger2002,
       author = {{Berger}, E.},
        title = "{Flaring up All Over-Radio Activity in Rapidly Rotating Late M and L Dwarfs}",
      journal = {\apj},
         year = 2002,
        month = jun,
       volume = {572},
       number = {1},
        pages = {503-513},
          doi = {10.1086/340301},
archivePrefix = {arXiv},
       eprint = {astro-ph/0111317},
 primaryClass = {astro-ph},
       adsurl = {https://ui.adsabs.harvard.edu/abs/2002ApJ...572..503B}
}

@ARTICLE{Burgasser_Putman2005,
       author = {{Burgasser}, Adam J. and {Putman}, Mary E.},
        title = "{Quiescent Radio Emission from Southern Late-Type M Dwarfs and a Spectacular Radio Flare from the M8 Dwarf DENIS 1048-3956}",
      journal = {\apj},
         year = 2005,
        month = jun,
       volume = {626},
       number = {1},
        pages = {486-497},
          doi = {10.1086/429788},
archivePrefix = {arXiv},
       eprint = {astro-ph/0502365},
 primaryClass = {astro-ph},
       adsurl = {https://ui.adsabs.harvard.edu/abs/2005ApJ...626..486B}
}

@ARTICLE{Kao2016,
       author = {{Kao}, Melodie M. and {Hallinan}, Gregg and {Pineda}, J. Sebastian and {Escala}, Ivanna and {Burgasser}, Adam and {Bourke}, Stephen and {Stevenson}, David},
        title = "{Auroral Radio Emission from Late L and T Dwarfs: A New Constraint on Dynamo Theory in the Substellar Regime}",
      journal = {\apj},
         year = 2016,
        month = feb,
       volume = {818},
       number = {1},
          eid = {24},
        pages = {24},
          doi = {10.3847/0004-637X/818/1/24},
archivePrefix = {arXiv},
       eprint = {1511.03661},
 primaryClass = {astro-ph.SR},
       adsurl = {https://ui.adsabs.harvard.edu/abs/2016ApJ...818...24K}
}

@ARTICLE{McLean2011,
       author = {{McLean}, M. and {Berger}, E. and {Irwin}, J. and {Forbrich}, J. and {Reiners}, A.},
        title = "{Periodic Radio Emission from the M7 Dwarf 2MASS J13142039+1320011: Implications for the Magnetic Field Topology}",
      journal = {\apj},
         year = 2011,
        month = nov,
       volume = {741},
       number = {1},
          eid = {27},
        pages = {27},
          doi = {10.1088/0004-637X/741/1/27},
archivePrefix = {arXiv},
       eprint = {1107.1516},
 primaryClass = {astro-ph.SR},
       adsurl = {https://ui.adsabs.harvard.edu/abs/2011ApJ...741...27M}
}

@ARTICLE{Forbrich2009,
       author = {{Forbrich}, Jan and {Berger}, Edo},
        title = "{The First VLBI Detection of an Ultracool Dwarf: Implications for the Detectability of Sub-Stellar Companions}",
      journal = {\apjl},
         year = 2009,
        month = dec,
       volume = {706},
       number = {2},
        pages = {L205-L209},
          doi = {10.1088/0004-637X/706/2/L205},
archivePrefix = {arXiv},
       eprint = {0910.1349},
 primaryClass = {astro-ph.SR},
       adsurl = {https://ui.adsabs.harvard.edu/abs/2009ApJ...706L.205F}
}

@ARTICLE{Lynch2015,
       author = {{Lynch}, C. and {Mutel}, R.~L. and {G{\"u}del}, M.},
        title = "{Wideband Dynamic Radio Spectra of Two Ultra-cool Dwarfs}",
      journal = {\apj},
         year = 2015,
        month = apr,
       volume = {802},
       number = {2},
          eid = {106},
        pages = {106},
          doi = {10.1088/0004-637X/802/2/106},
archivePrefix = {arXiv},
       eprint = {1405.3516},
 primaryClass = {astro-ph.SR},
       adsurl = {https://ui.adsabs.harvard.edu/abs/2015ApJ...802..106L}
}

@ARTICLE{Hallinan2007,
       author = {{Hallinan}, G. and {Bourke}, S. and {Lane}, C. and {Antonova}, A. and {Zavala}, R.~T. and {Brisken}, W.~F. and {Boyle}, R.~P. and {Vrba}, F.~J. and {Doyle}, J.~G. and {Golden}, A.},
        title = "{Periodic Bursts of Coherent Radio Emission from an Ultracool Dwarf}",
      journal = {\apjl},
         year = 2007,
        month = jul,
       volume = {663},
       number = {1},
        pages = {L25-L28},
          doi = {10.1086/519790},
archivePrefix = {arXiv},
       eprint = {0705.2054},
 primaryClass = {astro-ph},
       adsurl = {https://ui.adsabs.harvard.edu/abs/2007ApJ...663L..25H}
}

@ARTICLE{Llama2018,
       author = {{Llama}, Joe and {Jardine}, Moira M. and {Wood}, Kenneth and {Hallinan}, Gregg and {Morin}, Julien},
        title = "{Simulating Radio Emission from Low-mass Stars}",
      journal = {\apj},
         year = 2018,
        month = feb,
       volume = {854},
       number = {1},
          eid = {7},
        pages = {7},
          doi = {10.3847/1538-4357/aaa59f},
archivePrefix = {arXiv},
       eprint = {1801.01507},
 primaryClass = {astro-ph.SR},
       adsurl = {https://ui.adsabs.harvard.edu/abs/2018ApJ...854....7L}
}

@ARTICLE{Leone1993,
       author = {{Leone}, F. and {Umana}, G.},
        title = "{Periodic radio emission from the helium strong stars HD 37017 and sigma ORI E.}",
      journal = {\aap},
         year = 1993,
        month = feb,
       volume = {268},
        pages = {667-670},
       adsurl = {https://ui.adsabs.harvard.edu/abs/1993A&A...268..667L}
}

@ARTICLE{das2025b,
       author = {{Das}, Barnali and {Shultz}, Matt E. and {Pritchard}, Joshua and {Rose}, Kovi and {Driessen}, Laura Nicole and {Wang}, Yuanming and {Zic}, Andrew and {Murphy}, Tara and {Sivakoff}, Gregory},
        title = "{Discovery of Main-sequence Radio Pulse emitters from widefield sky surveys}",
      journal = {\pasa},
         year = 2025,
        month = jul,
       volume = {42},
          eid = {e110},
        pages = {e110},
          doi = {10.1017/pasa.2025.10036},
archivePrefix = {arXiv},
       eprint = {2505.07195},
 primaryClass = {astro-ph.SR},
       adsurl = {https://ui.adsabs.harvard.edu/abs/2025PASA...42..110D}
}

@ARTICLE{Guirado2025,
       author = {{Guirado}, J.~C. and {Climent}, J.~B. and {Bergasa}, J.~D. and {P{\'e}rez-Torres}, M.~A. and {Marcaide}, J.~M. and {Pe{\~n}a-Mo{\~n}ino}, L.},
        title = "{Main-oval Auroral Emission from a T6 Brown Dwarf: Observations, Modeling, and Astrometry}",
      journal = {\apj},
         year = 2025,
        month = jul,
       volume = {987},
       number = {1},
          eid = {7},
        pages = {7},
          doi = {10.3847/1538-4357/add5f3},
archivePrefix = {arXiv},
       eprint = {2505.04506},
 primaryClass = {astro-ph.SR},
       adsurl = {https://ui.adsabs.harvard.edu/abs/2025ApJ...987....7G}
}

@incollection{Curiel.1.2026.SKA, author = {Salvador Curiel and author2 and author3 and author4 and author5},title = {},year = {2026},publisher = {},note = {arXiv search: Report number AASKAII/Curiel01},booktitle = {Advancing Astrophysics with the SKA -- II (AASKAII)}}

\end{document}